\documentclass[letterpaper,12pt]{article}

\usepackage{t1enc}
\usepackage{times,exscale}
\usepackage{fancyheadings}
\usepackage{fullpage}
\usepackage{amsmath}
\usepackage{amsbsy}
\usepackage{amsfonts}
\usepackage{graphicx}
\usepackage{pstricks,psfrag}
\usepackage{subfigure}
\usepackage{natbib}
\usepackage{epsfig}

\newcommand{\mbs}[1]{\boldsymbol{#1}}
\newcommand{\mbb}[1]{\mathbb{#1}}

  \def\bF{{\mbs{F}}}
  \def\bI{{\mbs{I}}}
  
 \def\bN{{\mbs{N}}} 
\def\bP{{\mbs{P}}} \def\bQ{{\mbs{Q}}} \def\bR{{\mbs{R}}}
  \def\bU{{\mbs{U}}}

\def\ba{{\mbs{a}}}  
 \def\be{{\mbs{e}}}

\def\bm{{\mbs{m}}}  
  
 \def\bt{{\mbs{t}}} \def\bu{{\mbs{u}}}
  \def\bx{{\mbs{x}}}
\def\by{{\mbs{y}}} 

\def\CAN{Cu-Al-Ni }
\def\setR{\mathbb{R}}
\def\R{\setR}
\def\dps{\displaystyle}

\title{\bf{A Constrained Sequential-Lamination Algorithm
  for the Simulation of Sub-Grid Microstructure in Martensitic
  Materials}}

\author{
        {\bf{S.~Aubry, M.~Fago and M.~Ortiz}} \\
        Graduate Aeronautical Laboratories \\
        California Institute of Technology \\
        Pasadena, CA 91125, USA}

\date{February 27, 2002 \\
      Submitted to: {\it Computer Methods in Applied
      Mechanics and Engineering} \\
      Revised: October 10, 2002}

\begin{document}
\maketitle

\begin{abstract}

We present a practical algorithm for partially relaxing multiwell
energy densities such as pertain to materials undergoing
martensitic phase transitions. The algorithm is based on
sequential lamination, but the evolution of the microstructure
during a deformation process is required to satisfy a continuity
constraint, in the sense that the new microstructure should be
reachable from the preceding one by a combination of branching
and pruning operations. All microstructures generated by the
algorithm are in static and configurational equilibrium. Owing to
the continuity constrained imposed upon the microstructural
evolution, the predicted material behavior may be path-dependent
and exhibit hysteresis. In cases in which there is a strict
separation of micro and macrostructural lengthscales, the
proposed relaxation algorithm may effectively be integrated into
macroscopic finite-element calculations at the subgrid level. We
demonstrate this aspect of the algorithm by means of a numerical
example concerned with the indentation of an \CAN shape memory
alloy by a spherical indenter.
\end{abstract}

\begin{center}
\end{center}

\bigskip

\section{Introduction}

Materials often are capable of adopting a multiplicity of crystal structures,
or phases, the relative stability of which depends on temperature, the state
of stress, and other factors. Under conditions such that several phases are
energetically favorable, e.~g., at the transition temperature in martensitic
materials, materials are often found to develop microstructure in nature or in
the laboratory. A central problem in mechanics concerns the prediction of
these microstructures and their effect on the effective or macroscopic
behavior of materials, including such scaling properties and size effects as
may result from their formation and evolution. When martensitic materials are
modelled within the confines of nonlinear elasticity, the coexistence of
phases confers their strain-energy density function a multiwell structure
\cite{eriksen, chu.james, bhattacharya, li.luskin}.  The
corresponding boundary value problems are characterized by energy functions
which lack weak sequential lower-semicontinuity, and the energy-minimizing
deformation fields tend to develop fine microstructure \cite{ball.james1,
ball.james2, dacorogna, muller}.

There remains a need at present for efficient numerical methods for solving
macroscopic boundary-value problems while simultaneously accounting for
microstructure development at the microscale. One numerical strategy consists
of attempting a direct minimization of a suitably discretized energy function.
For instance, Tadmor {\it et al} \cite{tadmor.smith.bernstein.kaxiras} have
applied this approach to the simulation of nanoindentation in silicon. The
energy density is derived from the Stillinger-Weber potential by recourse to
the Cauchy-Born approximation, and accounts for five phases of silicon. The
energy functional is discretized by an application of the finite-element
method.  Tadmor {\it et al} \cite{tadmor.waghmare.smith.kaxiras} pioneering
calculations predict the formation of complex phase arrangements under the
indenter, and such experimentally observed features as an
insulator-to-conductor transition at a certain critical depth of indentation.

Despite these successes, direct energy minimization is not without
shortcomings. Thus, analysis has shown (see, e.~g., \cite{muller}
for a review) that the microstructures which most effectively
relax the energy may be exceedingly intricate and, consequently,
unlikely to be adequately resolved by a fixed numerical grid. As
a result, the computed microstructure is often coarse and biased
by the computational mesh, which inhibits---or entirely
suppresses---the development of many of the competing
microstructures. By virtue of these constraints, the numerical
solution is often caught up in a metastable local minimum which
may not accurately reflect the actual energetics and deformation
characteristics of the material.

In applications where there is a clear separation of micro and
macrostructural lengthscales, an alternative numerical strategy
is to use a suitably relaxed energy density in calculations
\cite{chipot.kinderlehrer, luskin,
bartels.carstensen.plechac.prohl, desimone, GovindjeeMiehe2001}.
In this approach, the original multiwell energy density is
replaced by its \emph{quasiconvex} envelop, i.~e., by the lowest
energy density achievable by the material through the development
of microstructure. Thus, the determination of the relaxed energy
density requires the evaluation of \emph{all} possible
microstructures compatible with a prescribed macroscopic
deformation. The resulting relaxed energy density is quasiconvex
\cite{muller}, and its minimizers, which represent the
macroscopic deformation fields, are devoid of microstructure and,
thus, more readily accessible to numerical methods. In essence,
the use of relaxed energy densities in macroscopic boundary value
problems constitutes a multiscale approach in which the
development of microstructure occurs---and is dealt with---at the
\emph{subgrid} level. The central problem in this approach is to
devise effective means of determining the relaxed energy density
and of integrating it into macroscopic calculations.

Unfortunately, no general algorithm for the determination of the
quasiconvex envelop of an arbitrary energy density is known at
present. A fallback strategy consists of the consideration of
\emph{special} microstructures only, inevitably resulting in a
partial relaxation of the energy density. For instance, attention
may be restricted to microstructures in the form of
\emph{sequential laminates} \cite{kohn91, luskin, ortiz.repetto,
ortiz.repetto.stainier, dolzmann, desimone}. The lowest energy
density achievable by the material through sequential lamination
is known as the \emph{rank-one} convexification of the energy
density. Many of the microstructures observed in shape-memory
alloys \cite{chu.james} and in ductile single crystals
\cite{ortiz.repetto} may be interpreted as instances of
sequential lamination, which suggests that the rank-one
convexification of the energy coincides---or closely
approximates---the relaxed energy for these materials. Ductile
single crystals furnish a notable example in which the rank-one
and the quasiconvex envelops are known to coincide exactly
\cite{aubry.bhattacharya.ortiz}.

In this paper we present a practical algorithm for partially
relaxing multiwell energy densities. The algorithm is based on
sequential lamination and, hence, at best it returns the rank-one
convexification of the energy density. Sequential lamination
constructions have been extensively used in both analysis and
computation \cite{kohn91, ball.james2, bhattacharya,
james.kinderlehrer, luskin, ortiz.repetto, Kruzik1998,
ArandaPedregal2001a, ArandaPedregal2001b}. All microstructures
generated by the algorithm are in static and configurational
equilibrium. Thus, we optimize all the interface orientations and
variant volume fractions, with the result that all configurational
forces and torques are in equilibrium. We additionally allow the
variants to be arbitrarily stressed and enforce traction
equilibrium across all interfaces.

The proposed lamination construction is \emph{constrained} in an
important respect: during a deformation process, we require that
every new microstructure be reachable from the preceding
microstructure along an admissible transition path. The mechanisms
by which microstructures are allowed to effect topological
transitions are: \emph{branching}, i.~e., the splitting of a
variant into a rank-one laminate; and \emph{pruning}, consisting
of the elimination of variants whose volume fraction reduces to
zero. Branching transitions are accepted provided that they reduce
the total energy, without consideration of energy barriers. By
repeated branching and pruning microstructures are allowed to
evolve along a deformation process. The \emph{continuation}
character of the algorithm furnishes a simple model of
metastability and hysteresis. Thus, successive microstructures are
required to be `close' to each other, which restrict the range of
microstructures accessible to the material at any given time. In
general, this restriction causes the microstructures to be
path-dependent and metastable, and the computed macroscopic
response may exhibit hysteresis.

The proposed relaxation algorithm may effectively be integrated
into macroscopic finite-element calculations at the subgrid
level. We demonstrate the performance and versatility of the
algorithm by means of a numerical example concerned with the
indentation of an \CAN shape memory alloy \cite{chu.james} by a
spherical indenter. The calculations illustrate the ability of
the algorithm to generate complex microstructures while
simultaneously delivering the macroscopic response of the
material. In particular, the algorithm results in force-depth of
indentation curves considerably softer than otherwise obtained by
direct energy minimization.

\section{Problem Formulation}

Let $\Omega \in \R^3$ be a bounded domain representing the
reference configuration of the material. Let $\by(x): \Omega \to
\R^3$ be the deformation and $\bF(x) = D \by(x)$ be the
corresponding deformation gradient. We denote the elastic energy
density at deformation gradient $\bF \in \R^{3 \times 3}$ by
$W(\bF)$. We require $W(\bF)$ to be material frame indifferent,
i.~e., to be such that $W(\bR\bF) = W(\bF)$, $\forall \bR \in
SO(3)$ and $\bF \in \R^{3\times3}$. In addition, the case of
primary interest here concerns materials such that $W(\bF)$ is not
quasiconvex. As a simple example, we may suppose that $W(\bF)$ has
the following special structure: Let $W_i(\bF)$, $i=1,\dots, M$
be quasiconvex energy densities (see, e.~g., \cite{dacorogna} for a
definition and discussion of quasiconvexity), representing the
energy wells of the material. Then
\begin{equation}\label{eq:Multiwell}
W(\bF) = \min_{m=0,\dots, M} W_m(\bF)
\end{equation}
i.~e., $W(\bF)$ is the lower envelop of the functions $W_m(\bF)$.

A common model of microstructure development in this class of
materials presumes that the microstructures of interest
correspond to low-energy configurations of the material, and that,
consequently, their essential structure may be ascertained by
investigating the absolute minimizers of the energy. However, the
energy functionals resulting from multiwell energy densities such
as (\ref{eq:Multiwell}) lack weak-sequential lower semicontinuity
and their infimum is not attained in general \cite{dacorogna}. The
standard remedy is to introduce the quasiconvex envelop
\begin{equation}
QW(\bF) = \frac{1}{|Q|} \inf_{\bu \in W^{1,\infty}_0(\Omega)}
\int_Q W(\bF + D \bu) dx
\end{equation}
of $W(\bF)$, or relaxed energy density. In this expression,
$W^{1, \infty}_0(\Omega)$ denotes the space of functions whose
distributional derivatives are essentially bounded and which
vanish on the boundary, and $Q$ is an arbitrary domain.
Physically, $QW(\bF)$ represents the lowest energy density
achievable by the material through the development of
microstructure. The macroscopic deformations of the solid are
then identified with the solutions of the relaxed problem:
\begin{equation}\label{Relaxed}
\inf_{\by\in X} \left\{ \int_\Omega [QW(D\by) - \mbs{f}\cdot \by ]
dx - \int_{\partial\Omega_2} \bar{\bt}\cdot\by\right\}
\end{equation}
where $X$ denotes some suitable solution space, $\mbs{f}$ is a
body force field, $\bt$ represents a distribution of tractions
over the traction boundary $\partial\Omega_2$, and the
deformation of the body is constrained by displacement boundary
conditions of the form:
\begin{equation}
\by = \bar{\by} \text{ on } \partial\Omega_1 = \partial\Omega -
\partial\Omega_2
\end{equation}
Thus, in this approach the effect of microstructure is built into
the relaxed energy $QW(\bF)$. The relaxed problem defined by
$QW(\bF)$ then determines the macroscopic deformation.

\begin{figure}[htbp]
  \begin{center}
    \psfrag{lambda1}{$\lambda_1$}
    \psfrag{lambda2}{$\lambda_2$}
    \psfrag{N1}{$\bN_1$}
    \psfrag{N2}{$\bN_2$}
    \includegraphics[width=8cm]{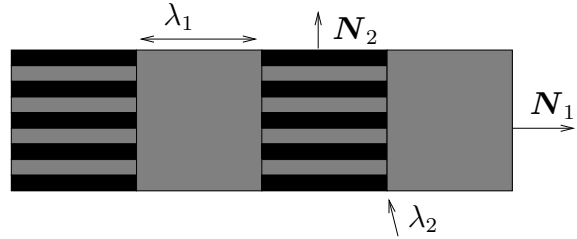}
  \end{center}
  \caption{Example of a rank-$2$ laminate. $\lambda_1$
  and $\lambda_2$ are the volume fractions corresponding to
  levels 1 and 2, respectively, and $\bN_1$ and $\bN_2$ are
  the corresponding unit normals.}
  \label{fig:lami}
\end{figure}

In executing this program the essential difficulty resides in the
determination of the relaxed energy $QW(\bF)$. As mentioned in the
introduction, no general algorithm for the determination of the
quasiconvex envelop of an arbitrary energy density is known at
present. A fallback strategy is to effect a partial relaxation of
the energy density by recourse to \emph{sequential lamination}
\cite{kohn91, luskin}, and the use of the resulting rank-one
convexification $RW(\bF)$ of $W(\bF)$ in lieu of $QW(\bF)$ in the
macroscopic variational problem (\ref{Relaxed}). We recall that
the rank-one convexification $RW(\bF)$ of $W(\bF)$ follows as the
limit \cite{KohnStrang1983, kohn.strang}
\begin{equation}
RW(\bF) = \lim_{k \to \infty } R_kW(\bF)
\end{equation}
where $R_0W(\bF) = W(\bF)$ and $R_kW(\bF)$ is defined recursively
as
\begin{equation}\label{eq:Rk}
\begin{split}
R_kW(\bF) &= \inf_{\lambda, \ba, \bN}\left\{ (1-\lambda)
R_{k-1}W(\bF-\lambda \ba \otimes \bN) +
\lambda R_{k-1}W(\bF+ (1-\lambda) \ba \otimes \bN), \right. \\
& \left. \lambda \in [0,1], \ba, \bN \in \R^3, |\bN| = 1\right\}
\quad k \geq 1
\end{split}
\end{equation}
In these expressions, $\lambda$ and $1-\lambda$ represent the
volume fractions of the $k$-level variants, $\bN$ is the unit
normal to the planar interface between the variants, and $\ba$ is
a vector (see Fig.~\ref{fig:lami}).

Unfortunately, a practical algorithm for the evaluation of the
rank-one convexification of general energy densities `on the fly'
does not appear to be available at present. Dolzmann
\cite{dolzmann} has advanced a method for the computation of the
exact rank-one convexification of an arbitrary energy density in
two dimensions. However, extensions of the method to three
dimensions do not appear to be available at present. In addition,
the method requires the {\it a priori} tabulation of $RW(\bF)$
over all of $\mbb{R}^{2\times 2}$, which limits its applicability
to large-scale computing.

An additional complication arises from the fact that the cyclic behavior of
martensitic materials often exhibits hysteresis. Under these conditions, the
response of the material is path-dependent and dissipative, and, therefore,
absolute energy minimization does not furnish a complete model of material
behavior. The modeling of hysteresis requires consideration of entire
deformation processes, rather than isolated states of deformation of the
material. A framework for the understanding of hysteresis may be constructed
by assuming that the evolution of microstructure is subject to a
\emph{continuity} requirement, namely, to the requirement that successive
microstructures be close to each other in some appropriate sense. This
constraint restricts the range of microstructures which the material may adopt
at any given time and thus results in metastable configurations. The
particular sequence of metastable configurations adopted by the material may
be path dependent, resulting in hysteresis. The connection between
metastability and hysteresis has been discussed by Ball, Chu and James
\cite{ball.chu.james}.

\section{A sequential lamination algorithm}

The problem which we address in the remainder of this paper
concerns the formulation of efficient algorithms for the
evaluation of $RW(\bF)$, and extensions thereof accounting for
kinetics and nonlocal effects, with specific focus on algorithms
which can be effectively integrated into large-scale macroscopic
simulations. We begin by reviewing basic properties of sequential
laminates for subsequent reference. More general treatments of
sequential lamination may be found in \cite{kohn.strang, kohn91,
bhattacharya91, bhattacharya92, pedregal, luskin}.

Uniform deformations may conventionally be categorized as
rank-zero laminates.  A rank-one laminate is a layered mixture of
two deformation gradients, $\bF^-$, $\bF^+ \in \R^{3\times3}$.
Compatibility of deformations then requires $\bF^\pm$ to be
rank-one connected, i.~e.,
\begin{equation}
\bF^+ - \bF^- = \ba \otimes \bN
\end{equation}
where $\ba \in \R^3$, and $\bN \in \R^3$, $|\bN|=1$, is the normal
to the interface between the two variants of deformation. Let
$\lambda^\pm$,
\begin{equation}
\lambda^- + \lambda^+ = 1, \quad \lambda^\pm \in [0, 1],
\end{equation}
denote the volume fractions of the variants. Then, the average or
macroscopic deformation follows as
\begin{equation}
\bF = \lambda^- \bF^- + \lambda^+ \bF^+
\end{equation}
If $\bF$ and $\{\ba, \lambda^\pm, \bN\}$ are known, then the
deformation in the variants is given by:
\begin{eqnarray}
  \label{eq:F+F-}
  \begin{array}{l}
    \bF^+ = \bF + \, \lambda^- \ba \otimes \bN\\
    \bF^- = \bF - \, \lambda^+ \ba \otimes \bN.
  \end{array}
\end{eqnarray}
and, thus, $\bF$ and $\{\ba, \lambda^\pm, \bN\}$ define a complete
set of--deformation and configurational--degrees of freedom for
the laminate. Following Kohn \cite{kohn91}, a laminate of
rank-$k$ is a layered mixture of two rank-$(k-1)$ laminates,
which affords an inductive definition of laminates of any rank.
As noted by Kohn \cite{kohn91}, the construction of sequential
laminates assumes a separation of scales: the length scale $l_k$
of the $k$th-rank layering satisfies $l_k\ll l_{k-1}$.

Sequential laminates have a binary-tree structure. Indeed, with
every sequential laminate we may associate a \emph{graph} $G$
such that: the nodes of $G$ consist of all the sub-laminates of
rank less or equal to the rank $k$ of the laminate; and joining
each sub-laminate of order $1\leq l\leq k$ with its two
constituent sub-laminates of order $l-1$. The root of the graph
is the entire laminate. Two sequential laminates will be said to
have the same \emph{structure} (alternatively, \emph{topology} or
\emph{layout}) if their graphs are identical. Evidently, having
the same structure defines an equivalence relation between
sequential laminates, and the set of all equivalence classes is in
one-to-one correspondence with the set ${\cal B}$ of binary trees.

\begin{figure}[htbp]
  \begin{center}
    \psfrag{F}{$\bF$}
    \psfrag{F-}{$\bF^-$}
    \psfrag{F+}{$\bF^+$}
    \psfrag{F-+}{$\bF^{- +}$}
    \psfrag{F--}{$\bF^{- -}$}
    \psfrag{L}{$\lambda^+$}
    \psfrag{L2}{$\lambda^-$}
    \psfrag{L3}{$\lambda^{- +}$}
    \psfrag{L4}{$\lambda^{--}$}
    \psfrag{nu}{$\nu_1$}
    \psfrag{nu2}{$\nu_2$}
    \includegraphics[width=10cm]{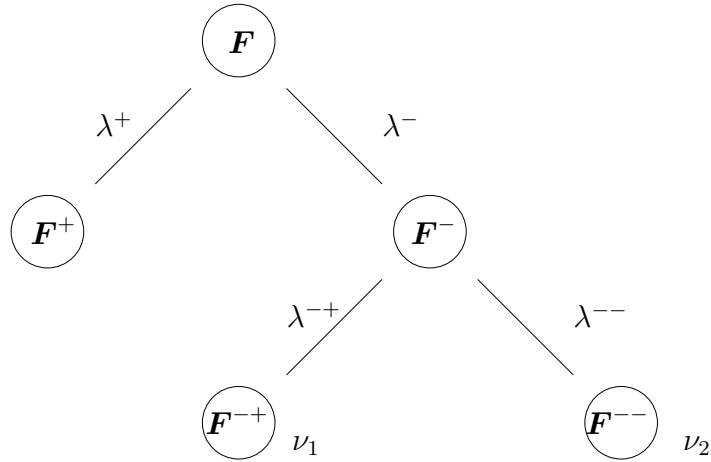}
  \end{center}
  \caption{\em Example of a rank-$2$ laminate. In this example,
    $\nu_1 = \lambda^-\lambda^{-+}$ and $\nu_2 =
    \lambda^-\lambda^{--}$.}
  \label{fig:tree}
\end{figure}

Let $i=1,\dots,n$ be an enumeration of the nodes of $G$. Then, to
each node $i$ we may associate a deformation $\bF_i$. The root
deformation is the average or macroscopic deformation $\bF$. Each
node in the tree has either two children or none at all. Nodes
with a common parent are called siblings. Nodes without children
are called leaves. Nodes which are not leaves are said to be
interior.  The deformations of the children of node $i$ will be
denoted $\bF_i^\pm$. Each generation of nodes is called a level.
The root occupies level $0$ of the tree.  The number of levels is
the rank $k$ of the tree.  Level $l$ contains at most $2^l$
nodes. The example in Fig.~\ref{fig:tree} represents a rank-two
laminate of order four. The three leaves of the tree are nodes
$\bF^+$, $\bF^{-+}$ and $\bF^{--}$. The interior nodes are $\bF$
and $\bF^-$. The children of, e.~g., node $\bF^-$ are nodes
$\bF^{-+}$ and $\bF^{--}$.

Compatibility demands that each pair of siblings be rank-one
connected, i.~e.,
\begin{equation}
\bF_i^+ - \bF_i^- = \ba_i \otimes \bN_i, \quad i \in {\cal I}_G
\end{equation}
where $\ba_i \in \R^3$, $\bN_i \in \R^3$, $|\bN|=1$, is the
normal to the interface between $\bF_i^-$ and $\bF_i^+$, and
${\cal I}_G$ denotes the set of all interior nodes. Let
$\lambda_i^\pm$,
\begin{equation}
\lambda_i^- + \lambda_i^+ = 1, \quad \lambda_i^\pm \in [0, 1],
\end{equation}
denote the volume fractions the variants $\bF_i^\pm$ within node
$i$. Then, the deformation of the parent variant is recovered in
the form
\begin{equation}
\bF_i = \lambda_i^- \bF_i^- + \lambda_i^+ \bF_i^+
\end{equation}
If $\bF_i$ and $\{\ba_i, \lambda_i^\pm, \bN_i\}$ are known for an
interior node $i$, then the deformation of its children is given
by:
\begin{eqnarray}
  \label{eq:SequentialF+F-}
  \begin{array}{l}
    \bF_i^+ =  \bF_i + \, \lambda_i^- \ba_i \otimes \bN_i\\
    \bF_i^- =  \bF_i - \, \lambda_i^+ \ba_i \otimes \bN_i.
  \end{array}
\end{eqnarray}
Therefore, $\bF$ and $\{\ba_i, \lambda_i^\pm, \bN_i, i \in {\cal
I}\}$ define a complete and independent set of degrees of freedom
for the laminate. A recursive algorithm for computing all the
variant deformations $\bF_i$, $i=1,\dots,n$ from $\bF$ and
$\{\ba_i, \lambda_i^\pm, \bN_i, i \in {\cal I}_G\}$ has been given
by \cite{ortiz.repetto.stainier}.

We shall also need the global volume fractions $\nu_l$ of all
leaves $l \in {\cal L}_G$, where ${\cal L}_G$ denotes the
collection of all leaves of $G$. These volume fractions are
obtained recursively from the relations:
\begin{equation}
\nu_i^\pm = \lambda_i^\pm \nu_i, \quad i \in {\cal I}_G
\end{equation}
with $\nu_{\rm root}=1$ for the entire laminate, and satisfy the
relation:
\begin{equation}
\sum_{l \in {\cal L}_G} \nu_l = 1
\end{equation}
Thus, $\nu_l$ represents the volume occupied by leaf $l$ as a
fraction of the entire laminate. The Young measure (e.~g.,
\cite{muller}) of the laminate consist of a convex combination
of atoms $\delta_{\bF_l}(\bF)$ with weights $\nu_l$, $l \in {\cal
L}$.

The average or macroscopic stress of the laminate may be
expressed in the form
\begin{equation}\label{eq:P}
\bP = \sum_{l \in {\cal L}_G} \nu_l \bP_l
\end{equation}
where
\begin{equation}
\bP_l = W,_{\bF}(\bF_l), \quad l \in {\cal L}_G
\end{equation}
are the first Piola-Kirchhoff stresses in the leaves. The average
stress may be computed by recursively applying the averaging
relation:
\begin{equation}
\bP_i = \lambda^- \bP_i^- + \lambda^+ \bP_i^+, \quad i \in {\cal
I}
\end{equation}
starting from the leaves of the laminate. A recursive algorithm
for computing the average stress $\bP$ from $\{\bP_l, l \in {\cal
L} \}$ and $\{\lambda_i^\pm, i \in {\cal I}_G\}$ has been given by
\cite{ortiz.repetto.stainier}.

\subsection{Microstructural equilibrium}

We begin by investigating the mechanical and configurational
equilibrium of sequential laminates of given structure. Thus, we
consider an elastic material of strain-energy density $W(\bF)$
subject to a prescribed macroscopic deformation $\bF \in
\mbb{R}^{3\times 3}$. In addition, we consider all sequential
laminates with given graph $G$.

The equilibrium configurations of the laminate then follow as the
solutions of the constrained minimization problem:
\begin{equation}
GW(\bF) = \inf_{\{\ba_i, \lambda_i^\pm, \bN_i, i \in {\cal I}_G\}}
\quad \sum_{l \in {\cal L}_G} \nu_l W(\bF_l) \label{eq:minW}
\end{equation}
\vskip -0.250in
\begin{eqnarray}
& \lambda_i^\pm \in [0 \mbox{, } 1], \quad i \in {\cal I}_G &
\label{eq:minW2} \\
& | \bN_i | = 1, \quad i \in {\cal I}_G & \label{eq:minW3}
\end{eqnarray}
where the $\bF_l$ are obtained from the recursive relations
(\ref{eq:SequentialF+F-}). The effective or macroscopic energy 
of the laminate is $GW(\bF)$. If $W(\bF)$ is quasiconvex then it
 necessarily follows that $\ba_i = {\bf 0}$, $\forall i \in {\cal
I}_G$, and $GW(\bF) = W(\bF)$.

It is interesting to verify that the solutions of the minimization
problem (\ref{eq:minW}) are in both force and configurational
equilibrium. Thus, assuming sufficient smoothness, the
stationarity of the energy with respect to all deformation jump
amplitudes yields the traction equilibrium equations:
\begin{equation}\label{eq:ForceEq}
(\bP_i^+ - \bP_i^-) \cdot \bN_i = {\bf 0}, \quad i \in {\cal I}_G
\end{equation}
Stationarity with respect to all normal vectors yields the
configurational-torque equilibrium equations:
\begin{equation}
[\ba_i \cdot (\bP_i^+ - \bP_i^-)] \times \bN_i = {\bf 0}, \quad i
\in {\cal I}_G
\end{equation}
Finally, stationarity with respect to all volume fractions yields
the configurational-force equilibrium equations:
\begin{equation}\label{eq:CEq}
f_i = (W_i^+ - W_i^-) - (\lambda_i^+\bP_i^+ + \lambda_i^-\bP_i^-)
\cdot (\ba_i \otimes \bN_i) = 0, \quad i \in {\cal I}_G
\end{equation}
where $f_i$ is the configurational force which drives interfacial
motion \cite{bhattacharya}. It bears emphasis that the leaf
deformations $\bF_l$ may in general be arbitrarily away from the
minima of $W(\bF)$, and thus the equilibrium equations
(\ref{eq:ForceEq}) must be carefully enforced. In addition, the
minimization (\ref{eq:minW}) has the effect of \emph{optimizing}
the volume fractions of all variants and the corresponding
interface orientations. If all the preceding stationarity
conditions are satisfied, then it is readily verified that the
average or macroscopic deformation (\ref{eq:P}) is recovered as
\begin{equation}
\bP = GW,_{\bF}(\bF)
\end{equation}
which shows that $GW(\bF)$ indeed supplies a potential for the
average or macroscopic stress of the laminate.

The case in which the energy density $W(\bF)$ possesses the
multiwell structure (\ref{eq:Multiwell}) merits special mention.
In this case, the minimization problem
(\ref{eq:minW}-\ref{eq:minW3}) may be written in the form:
\begin{equation}
GW(\bF) = \inf_{
\begin{array}{c}
\{\ba_i, \lambda_i^\pm, \bN_i, \ i \in {\cal I}_G \} \\
\{ m_l \in \{1, \dots, M\}, \ l \in {\cal L}_G \} \\
\end{array}}
\quad \sum_{l \in {\cal L}_G } \nu_l W_{m_l}(\bF_l)
\label{eq:MWminW}
\end{equation}
\vskip -0.15in
\begin{eqnarray}
& \lambda_i^\pm \in [0 \mbox{, } 1], \quad i \in {\cal I}_G &
\label{eq:MWminW2} \\
& | \bN_i | = 1, \quad i \in {\cal I}_G & \label{eq:MWminW3}
\end{eqnarray}
where $\bm = \{m_l \in \{1, \dots, M \}, \ l \in {\cal L}_G \}$
denotes the collection of wells which are active in each of the
leaves. This problem may be conveniently decomposed into two
steps: a first step involving energy minimization for a
prescribed distribution of active wells, namely,
\begin{equation}
G_{\bm}W(\bF) = \inf_{ \{\ba_i, \lambda_i^\pm, \bN_i, \ i \in
{\cal I}_G \} } \quad \sum_{l \in {\cal L}_G } \nu_l
W_{m_l}(\bF_l) \label{eq:MW2minW}
\end{equation}
\vskip -0.150in
\begin{eqnarray}
& \lambda_i^\pm \in [0 \mbox{, } 1], \quad i \in {\cal I}_G &
\label{eq:MW2minW2} \\
& | \bN_i | = 1, \quad i \in {\cal I}_G & \label{eq:MW2minW3}
\end{eqnarray}
followed by the optimization of the active wells, i.~e.,
\begin{equation}
GW(\bF) = \inf_{\{ m_l \in \{1, \dots, M \}, \ l \in {\cal L}_G
\}} G_{\bm}W(\bF)
\end{equation}

It should be carefully noted that the minimizers of problem
(\ref{eq:minW}-\ref{eq:minW3}) may be such that one or more of
the volume fractions $\lambda_i^\pm$ take the limiting values of
$0$ or $1$. We shall say that a graph $G$ is \emph{stable} with
respect to a macroscopic deformation $\bF$ if at least one
minimizer of (\ref{eq:minW}-\ref{eq:minW3}) is such that
\begin{equation}
\lambda^\pm_i \in (0 \mbox{, } 1), \quad \forall i \in {\cal I}_G
\end{equation}
and we shall say that the graph is \emph{unstable} or
\emph{critical} otherwise. The presence of sub-trees of zero
volume in an unstable graph is an indication that the graph is not
`right' for the macroscopic deformation $\bF$, i.~e., the graph
is unable to support a nontrivial microstructure consistent with
$\bF$. Unstable graphs are mathematically contrived and physically
inadmissible, and, as such, should be ruled out by some
appropriate means. This exclusion may be accomplished, e.~g., by
the simple device of assigning the offending solutions an infinite
energy, which effectively rules them out from consideration; or
by defining solutions modulo \emph{null} sub-trees, i.~e.,
sub-trees of vanishing volume. In the present approach, we choose
to integrate the exclusion of null sub-trees into the dynamics by
which microstructures are evolved, as discussed next.

\subsection{Microstructural evolution}

The problem (\ref{eq:minW}-\ref{eq:minW3}) may be regarded as a
partial rank-one convexification of $W(\bF)$ obtained by
prescribing the graph $G$ of the test laminates. The full
rank-one convexification follows from the consideration of all
possible graphs, i.~e.,
\begin{equation}\label{eq:RWGW}
RW(\bF) = \inf_{G \in {\cal B}} GW(\bF)
\end{equation}
where, as before, ${\cal B}$ is the set of all binary trees. In
the particular case of energy densities of the form
(\ref{eq:Multiwell}), we alternatively have
\begin{equation}\label{eq:RWGW2}
RW(\bF) = \inf_{ \begin{array}{c} G \in {\cal B} \\
\{ m_l \in \{1, \dots, M \}, \ l \in {\cal L}_G \} \end{array} }
G_{\bm}W(\bF)
\end{equation}
It is clear that this problem exhibits combinatorial complexity as
the rank of the test laminates increases, which makes a direct
evaluation of (\ref{eq:RWGW}) or (\ref{eq:RWGW2}) infeasible in
general.

Problems of combinatorial complexity arise in other areas of
mathematical physics, such as structural optimization and
statistical mechanics. Common approaches to the solution of these
problems are to restrict the search to the most `important' states
within phase space, or importance sampling; or to restrict access
to phase space by the introduction of some form of dynamics. In
this latter approach the states at which the system is evaluated
form a sequence, or `chain', and the next state to be considered
is determined from the previous states in the chain. If, for
instance, only the previous state is involved in the selection of
the new state, a Markov chain is obtained. In problems of energy
minimization, a common strategy is to randomly `flip' the system
and accept the flip with probability one if the energy is
reduced, and with a small probability if the energy is increased.

In other cases, the system possesses some natural dynamics which
may be exploited for computational purposes. A natural dynamics
for problem (\ref{eq:RWGW}) may be introduced as follows.
Evidently, the relevant phase space for this problem is ${\cal
B}$, the set of all binary trees, and the aim is to define a flow
$G(t)$ in this phase space describing the evolution of the
microstructure along a \emph{deformation processes} $\bF(t)$. Here
and subsequently, the real variable $t \geq 0$ denotes time. A
natural dynamics for $G(t)$ is set by the following conditions:

\begin{enumerate}

\item $G(t)$ must be stable with respect to $\bF(t)$.

\item $G(t)$ must be accessible from $G(t^-)$ through a
physically admissible transition.

\end{enumerate}

\noindent The first condition excludes laminates containing null
subtrees, i.~e., subtrees of zero volume. The second criterion
may be regarded as a set of rules for microstructural
\emph{refinement} and \emph{unrefinement}.

In order to render these criteria in more concrete terms, we
adopt an incremental viewpoint and seek to sample the
microstructure at discrete times $t_0 = 0$, $\dots$, $t_n$,
$t_{n+1}$, $\dots$. Suppose that the microstructure is known at
time $t_n$ and we are given a new macroscopic deformation
$\bF_{n+1} = \bF(t_{n+1})$. In particular, let $G_n$ be the graph
of the microstructure at time $t_n$. We consider two classes of
admissible transitions by which a new structure $G_{n+1}$ may be
reached from $G_n$:

\begin{enumerate}

\item The elimination of null subtrees from the graph of the
laminate, or \emph{pruning}.

\item The splitting of leaves, or \emph{branching}.

\end{enumerate}

\noindent Specifically, we refer to as branching the process by
which a leaf is replaced by a simple laminate. The criterion that
we adopt for accepting or rejecting a branching event is simple
energy minimization. Thus, let $l \in {\cal L}_{G_n}$ be a leaf
in the microstructure at time $t_n$, and let $\bF_l^n$ be the
corresponding deformation. The energetic `driving force' for
branching of the leaf $l$ may be identified with
\begin{equation}\label{BranchingDrivingForce}
f_l^n = W(\bF_l^n) - R_1W(\bF_l^n)
\end{equation}
where $R_1W(\bF)$ is given by (\ref{eq:Rk}). We simply accept or
reject the branching of the leaf $l \in {\cal L}_{G_n}$ according
as to whether $f_l^n > 0$ or $f_l^n \leq 0$, respectively.

In the particular case in which $W(\bF)$ is of the form
(\ref{eq:Multiwell}), the evaluation of $R_1W(\bF)$ may be
effected by considering all pairs of well energy densities, one
for each variant. However, since the well energy densities
$W_m(\bF)$ are assumed to be quasiconvex and we rule out variants
of zero volume, we may exclude from consideration the cases in
which both variants of the laminate are in the same well. The
evaluation of $R_1W(\bF)$ is thus reduced to the consideration of
all distinct pairs of well energy densities.

The precise sequence of steps followed in calculations are as
follows:

\begin{enumerate}

\item {Initialization:} Input $\bF_{n+1}$, set $G_{n+1} = G_n$.

\item {Equilibrium:} Equilibrate laminate with $G_{n+1}$ held
fixed.

\item {Evolution:}

\begin{enumerate}

\item Are there null subtrees?

\begin{enumerate}

\item {\tt YES:} Prune all null subtrees, {\tt GOTO} (2).

\item {\tt NO:} Continue.

\end{enumerate}

\item Compute all driving forces for branching $\{f_l^{n+1}, \ l \in
{\cal L}_{G_{n+1}} \}$. Let $f_{l_{\rm max}}^{n+1} = \max_{l \in
{\cal L}_{G_{n+1}}} f_l^{n+1}$. Is $f_{l_{\rm max}}^{n+1} > 0$?

\begin{enumerate}

\item {\tt YES:} Branch leaf $l_{\rm max}$, {\tt GOTO} (2).

\item {\tt NO:} EXIT.

\end{enumerate}

\end{enumerate}

\end{enumerate}

Several remarks are in order. The procedure just described may be
regarded as a process of \emph{continuation}, where the new
microstructure is required to be close to the existing one in the
sense just described. Evidently, since we restrict the class of
microstructures which may arise at the end of each time step,
there is no guarantee that this continuation procedure deliver
the solution of (\ref{eq:RWGW}) for all $t \geq 0$. However,
metastability plays an important role in many systems of
interest, and the failure to deliver the absolute rank-one
convexification at all times is not of grave concern in these
cases. Indeed, the continuation procedure described above may be
regarded as a simple model of metastability.

In this regard, several improvements of the model immediately
suggest themselves. Thus, the branching criterion employed in the
foregoing simply rules out branching in the presence of an
intervening energy barrier, no matter how small, separating the
initial and final states. An improvement over this model would be
to allow, with some probability, for transitions requiring an
energy barrier to be overcome, e.~g., in the spirit of transition
state theory and kinetic Monte Carlo methods. However, the
implementation of this approach would require a careful and
detailed identification of all the paths by which branching may
take place, a development which appears not to have been
undertaken to date.

\section{Illustrative examples}

As a first illustration of the sequential lamination algorithm
presented in the foregoing, we apply it to a simple model of a
\CAN shape-memory alloy, a material which has been extensively
investigated in the literature (cf \cite{ball.james1, ball.james2,
chu.james, luskin}, and references therein). Photomicrographs
taken from the experiments of Chu and James \cite{chu.james}
(also reported by \cite{luskin}) reveal sharp laminated
microstructures, often of rank two or higher. In order to
exercise the algorithm, in the examples that follow we simply take
the material through a prescribed macroscopic deformation path.

\subsection{Material model}

\CAN undergoes a cubic to orthorhombic martensitic transformation
at around room temperature and has, therefore, six variants in the
martensitic phase. The deformation undergone by the material in
transforming from austenite to an unstressed variant of
martensite may be described by a stretch tensor $\bU_m$,
$m=1,\dots 6$. For \CAN, these are \cite{chu.james, chu}:
\begin{equation}
\bU_1 =  \left(
{\begin{array}{ccc}
    \zeta & 0 & 0 \\
    0 & \xi  & \eta  \\
    0 & \eta  & \xi
  \end{array}}
\right)
\mbox{, }
\quad
\bU_2 =  \left(
  {\begin{array}{ccc}
      \zeta  & 0 & 0 \\
      0 & \xi  &  - \eta  \\
      0 &  - \eta  & \xi
    \end{array}}
\right)
\mbox{, }
\quad
\bU_3 =  \left(
  {\begin{array}{ccc}
      \xi  & 0 & \eta  \\
      0 & \zeta  & 0 \\
      \eta  & 0 & \xi
    \end{array}}
\right)
\end{equation}
\begin{equation}
\bU_4 =  \left(
  {\begin{array}{ccc}
      \xi  & 0 &  - \eta  \\
      0 & \zeta  & 0 \\
      - \eta  & 0 & \xi
    \end{array}}
\right)
\mbox{, }
\quad
\bU_5 =  \left(
  {\begin{array}{ccc}
      \xi  & \eta  & 0 \\
      \eta  & \xi  & 0 \\
      0 & 0 & \zeta
    \end{array}}
\right)
\mbox{, }
\quad
\bU_6 =  \left(
  {\begin{array}{ccc}
      \xi  &  - \eta  & 0 \\
      - \eta  & \xi  & 0 \\
      0 & 0 & \zeta
    \end{array}}
\right)
\mbox{.}
\end{equation}
where $\xi = 1.0425$, $\eta = 0.0194$ and $\zeta= 0.9178$, and
all components are referred to the cubic axes of the austenitic
phase. For purposes of illustration of the sequential lamination
algorithm we adopt a simple energy density of the form
(\ref{eq:Multiwell}), with well energy densities
\begin{equation}\label{eq:WCANAustenite}
  W_0(\bF) = \frac{1}{2}
  ( \bF^T \bF - \bI )  \, {{\mbb C}}_0 \,
  ( \bF^T \bF - \bI )
\end{equation}
for the austenitic phase, and
\begin{equation}\label{eq:WCAN}
  W_m(\bF) = \frac{1}{2}
  [ ( \bF \bU_m^{-1})^T ( \bF \bU_m^{-1}) - \bI ]  \, {{\mbb C}}_m \,
  [ ( \bF \bU_m^{-1})^T ( \bF \bU_m^{-1}) - \bI ], \quad m = 1,
  \dots, 6
\end{equation}
for the martensitic phases. In these expressions ${{\mbb C}}_m$,
$m=0, \dots, M$, are the elastic moduli at the bottom of the
variants. These are (in MPa):
\begin{equation}
{{\mbb C}}_0 = \left(
\begin{array}{cccccc}
C_{11} & C_{12} & C_{12} & 0     & 0    & 0\\
C_{12} & C_{11} & C_{12} & 0     & 0    & 0\\
C_{12} & C_{12} & C_{11} & 0     & 0    & 0\\
0    & 0    & 0    & C_{33}  & 0    & 0\\
0    & 0    & 0    & 0     & C_{33} & 0\\
0    & 0    & 0    & 0     & 0    & C_{33}
\end{array}
\right)
\end{equation}
where $C_{11} = 141.76$, $C_{12} = 126.24$ and $C_{33} =97$ and
\begin{equation}
{{\mbb C}}_1 = \left(
\begin{array}{cccccc}
C_{11} & C_{12} & C_{13} & 0   & 0    & 0 \\
C_{12} & C_{22} & C_{23} & 0   & 0    & 0 \\
C_{13} & C_{23} & C_{33} & 0   & 0    & 0 \\
0    & 0    & 0    & C_{44} & 0    & 0\\
0    & 0    & 0    & 0    & C_{55} & 0\\
0    & 0    & 0    & 0    & 0    & C_{66}
\end{array}
\right)
\end{equation}
where $C_{11} = 189$, $C_{22} = 141$ and
$C_{33} = 205$, $C_{12} = 124$, $C_{13} = 45.5$,
$C_{23} = 115$, $C_{44} = 54.9$, $C_{55} = 19.7$, $C_{66} = 62.6$ and
with the moduli of the remaining martensitic variants following by
symmetry. The energy density defined by (\ref{eq:Multiwell}) and
(\ref{eq:WCAN}) is material frame indifferent,  results in
stress-free states at the bottoms of all wells, i.~e., at
$\bF=\bI$ and $\bF=\bR\bU_m$, $\bR\in SO(3)$, $m=1,\dots, 6$,
assigns equal energy density to all unstressed variants, and
exhibits all the requisite material symmetries.

\subsection{Optimization}

In the examples presented here, and in the finite element
calculations presented in the following section, problem
(\ref{eq:minW}-\ref{eq:minW3}) is solved using Spellucci's
sequential quadratic programming (SQP) algorithm for constrained
minimization. The SQP algorithm is an interative procedure which
requires an initial guess in order to start the iteration. In
calculations we begin by setting the initial values of $\{\ba_i,
\lambda_i^\pm, \bN_i, i \in {\cal I}_{G_{n+1}}\}$ at time
$t_{n+1}$ equal to the converged values at time $t_n$. The main
issue arises in evaluating possible branching events, as in this
case new interfaces arise for which no previous geometrical
information exists. The selection of initial guesses for $\ba_i$
and $\lambda_i^\pm $, $i \in {\cal I}_G$, offers no difficulty,
and we simply set $\ba_i = {\bf 0}$ and $\lambda_i^\pm = 0$ or 1.
The choice of the initial value of the new interface normals
$\bN_i$ requires more care since it strongly biases the resulting
microstructure.

We have investigated two ways of initializing the normals $\bN_i$
arising during branching. A first approach is based on sampling
the unit sphere uniformly. Thus, we simply select initial values
of $\bN_i$ uniformly distributed over the unit sphere with some
prespecified density and select the solution which results in the
least energy. If two or more branched configurations possess the
same energy, we select one at random. This exhaustive search
approach is effective but costly owing to the large number of
cases which need to be considered.

The second approach consists of priming the iteration using an
initial guess derived from Ball and James  \cite{ball.james1,
ball.james2} constrained theory for shape-memory materials. In
this theory, the elastic moduli are presumed large compared to the
transformation stresses, so that the geometry of the laminate can
be obtained, to a first approximation, directly from the
transformation strains. Conveniently, all resulting twinning
relations between every distinct pair of martensitic wells can be
tabulated beforehand. For \CAN this tabulation has been carried
out by Bhattacharya, Li and Luskin \cite{bhattacharya.li.luskin}.
As expected from general theory, the only non-trivial rank-one
connections take place between distinct variants of martensite.
Specifically, we seek $\bQ \in SO(3)$, $\ba, \bN \in {{\mbb
R}}^3$, $|\bN| = 1$, such that
\begin{equation}
  \label{eq:rel}
  \bQ \bU_m = \bU_n + \ba \otimes \bN, \quad m, n = 1, \dots, 6, \
  m \neq n
\end{equation}
There are two solutions to this equation. Let $\widetilde{\be} =
\dps\be/\dps|\be|$, with the vector $\be$ as in Table~\ref{tb:e}.
The first solution is
\begin{equation}
  \ba = 2 \left(
    \frac{\bU_n^{-1}\widetilde{\be}}{|\bU_n^{-1}\widetilde{\be}|^2}  -
    \bU_n \widetilde{\be} \right), \quad
  \bN = \widetilde{\be}, \quad n = 1, \dots, 6, \ n \neq m
\end{equation}
and the second solution is
\begin{equation}
\ba =  C \bU_n \widetilde{\be}, \quad  \bN =  \frac{2}{C} \left(
\widetilde{\be} -
\frac{\bU_n^2\widetilde{\be}}{|\bU_n^2\widetilde{\be}|} \right),
\quad C = 2 \left|\widetilde{\be} -
\frac{\bU_n^2\widetilde{\be}}{|\bU_n^2\widetilde{\be}|} \right|,
\quad n = 1, \dots, 6, \ n \neq m
\end{equation}
\vskip 0.2cm
\begin{table}
  \centerline{
    \begin{tabular}{| c | c | c | c | c | c | c |}
      \hline wells &  $1$ & $2$ & $3$ & $4$ & $5$ & $6$ \\
      \hline $1$ &  & $\be_3$ & $\be_1 -\be_2$  & $\be_1 + \be_2$ & $\be_1 -
      \be_3$ &  $\be_1 + \be_3$\\
      \hline $2$ & $\be_2$ && $\be_1 + \be_2$  & $\be_1 - \be_2$  & $\be_1 +
      \be_3$ &  $\be_1 - \be_3$ \\
      \hline $3$ & $\be_1 -\be_2$  & $\be_1+\be_2$ & & $\be_3$ & $\be_2 -
      \be_3 $ &  $\be_2 + \be_3$  \\
      \hline $4$ & $\be_1+ \be_2$  & $\be_1-\be_2$ & $\be_3$ & & $\be_2 +
      \be_3 $&  $\be_2 - \be_3 $ \\
      \hline $5$ & $\be_1 - \be_3$  & $\be_1 +\be_3$ & $\be_2 - \be_3$ &
      $\be_2 + \be_3$ & & $\be_2$   \\
      \hline $6$ & $\be_1 + \be_3$ & $\be_1 -\be_3$  & $\be_2 + \be_3$ &
      $\be_2 -\be_3$ & $\be_1$ & \\
      \hline
    \end{tabular}
    }
  \caption{Vector $\be$ arising in the twinning relations for \CAN
  \cite{bhattacharya.li.luskin}. The vectors $\{\be_1, \be_2,
  \be_3\}$ correspond to the cube directions in the austenite phase.}
  \label{tb:e}
\end{table}
\noindent These results from the constrained theory may
conveniently be used to start a branching calculation when the
energy density is of the form (\ref{eq:Multiwell}). In this case,
the branching calculation entails the consideration of every pair
of well energy densities $W_m(\bF)$ and $W_n(\bF)$, $m, n = 1,
\dots, 6$, $m \neq n$, in the new leaves. The attendant iteration
may then be started from the constrained solutions $\{\ba, \bN\}$
described above, with the initial value of $\lambda^\pm$
determined by means of a line search.

\begin{figure}[htbp]
    \begin{center}
    \subfigure[]{\includegraphics[width=2.5in]{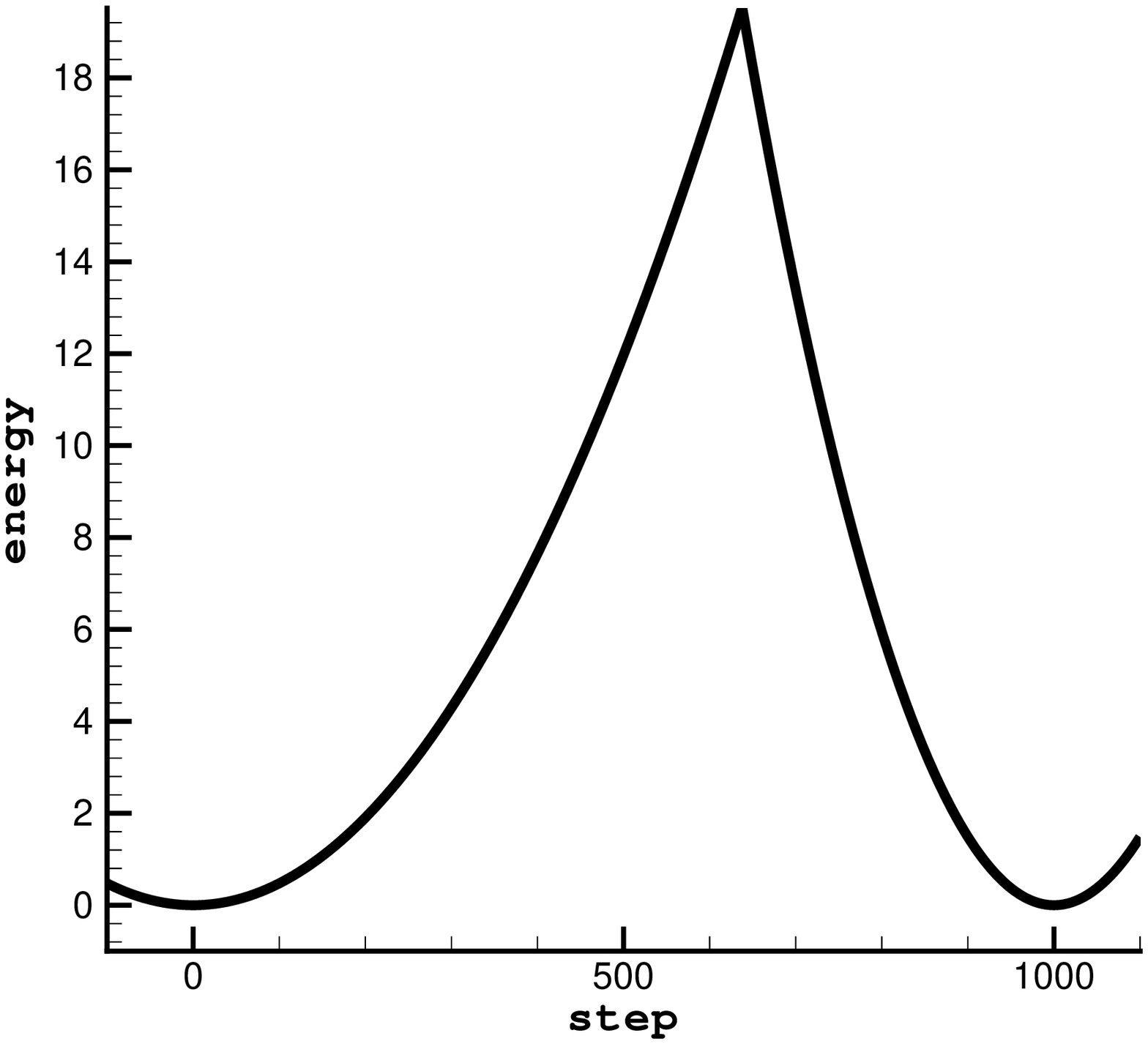}}
    \subfigure[]{\includegraphics[width=2.5in]{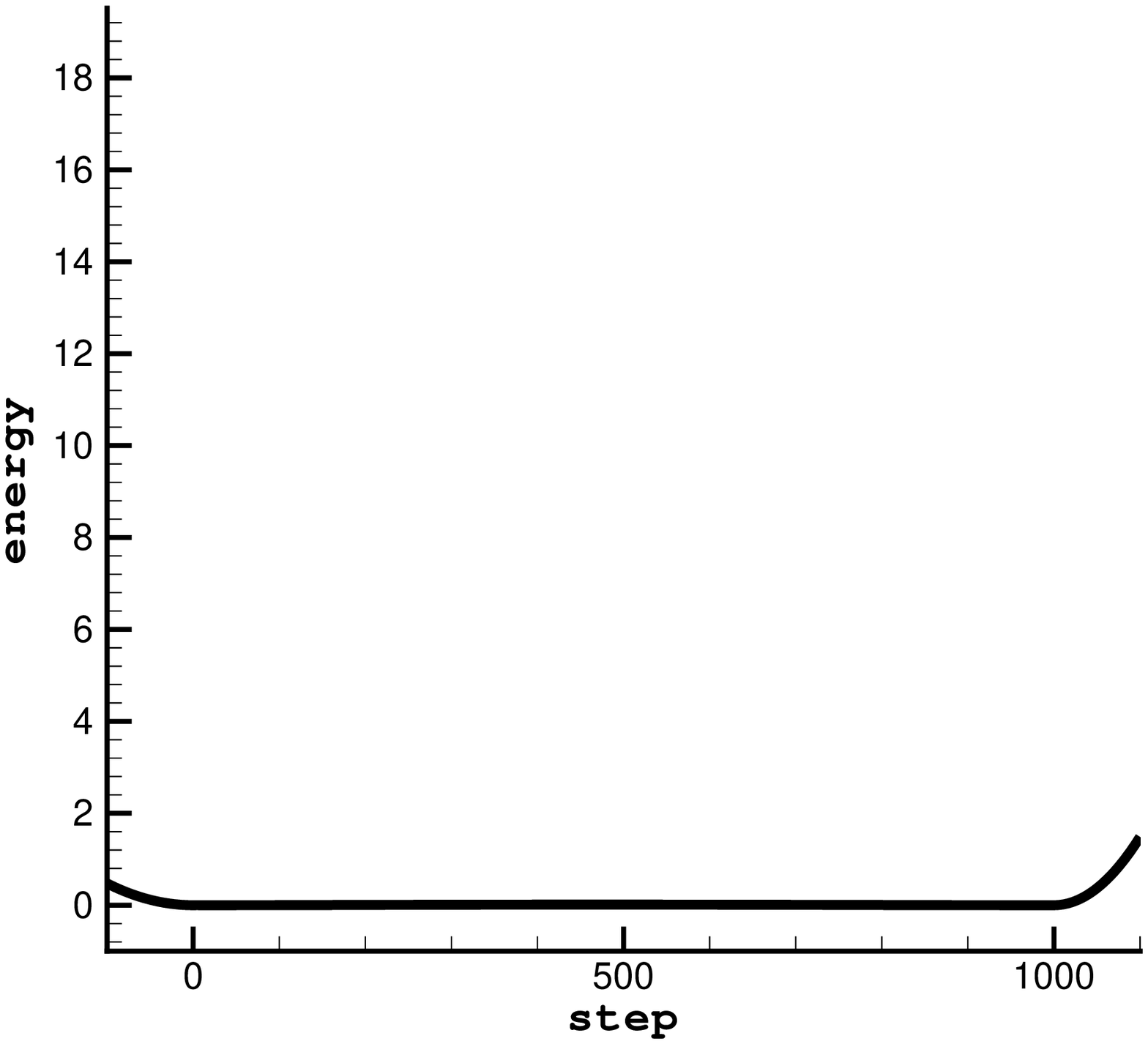}}
    \subfigure[]{\includegraphics[width=2.5in]{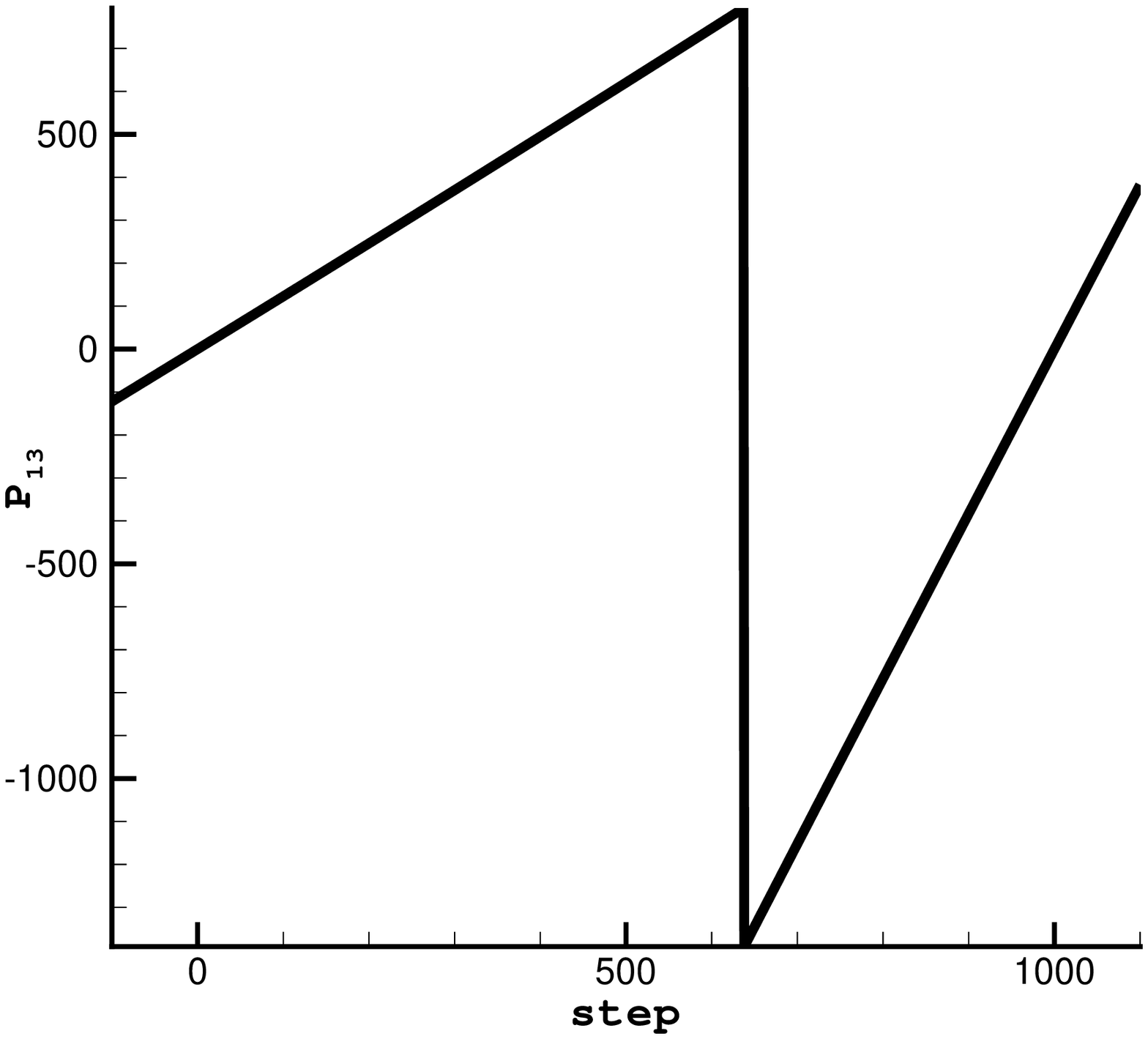}}
    \subfigure[]{\includegraphics[width=2.5in]{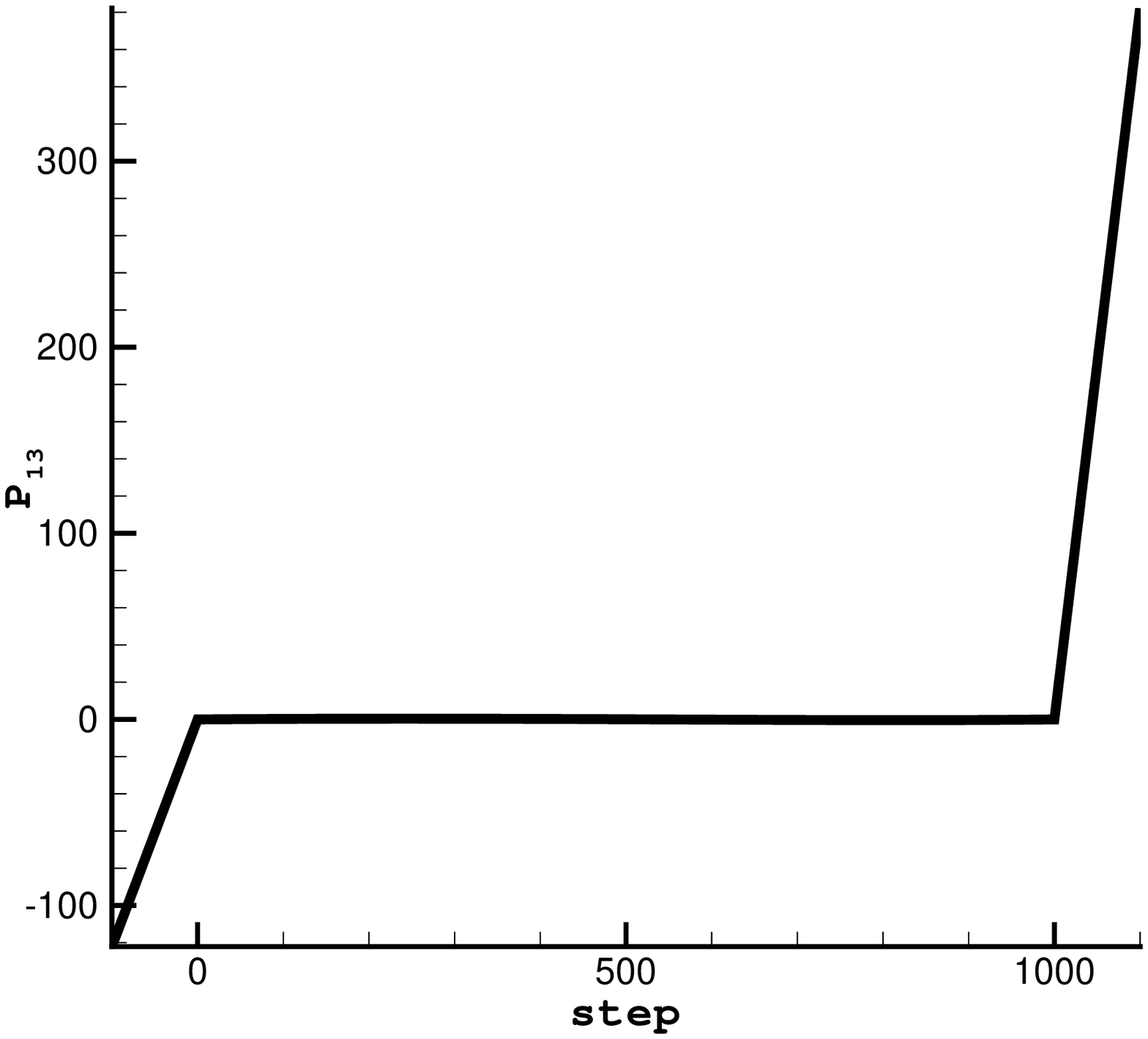}}
    \subfigure[]{\includegraphics[width=2.5in]{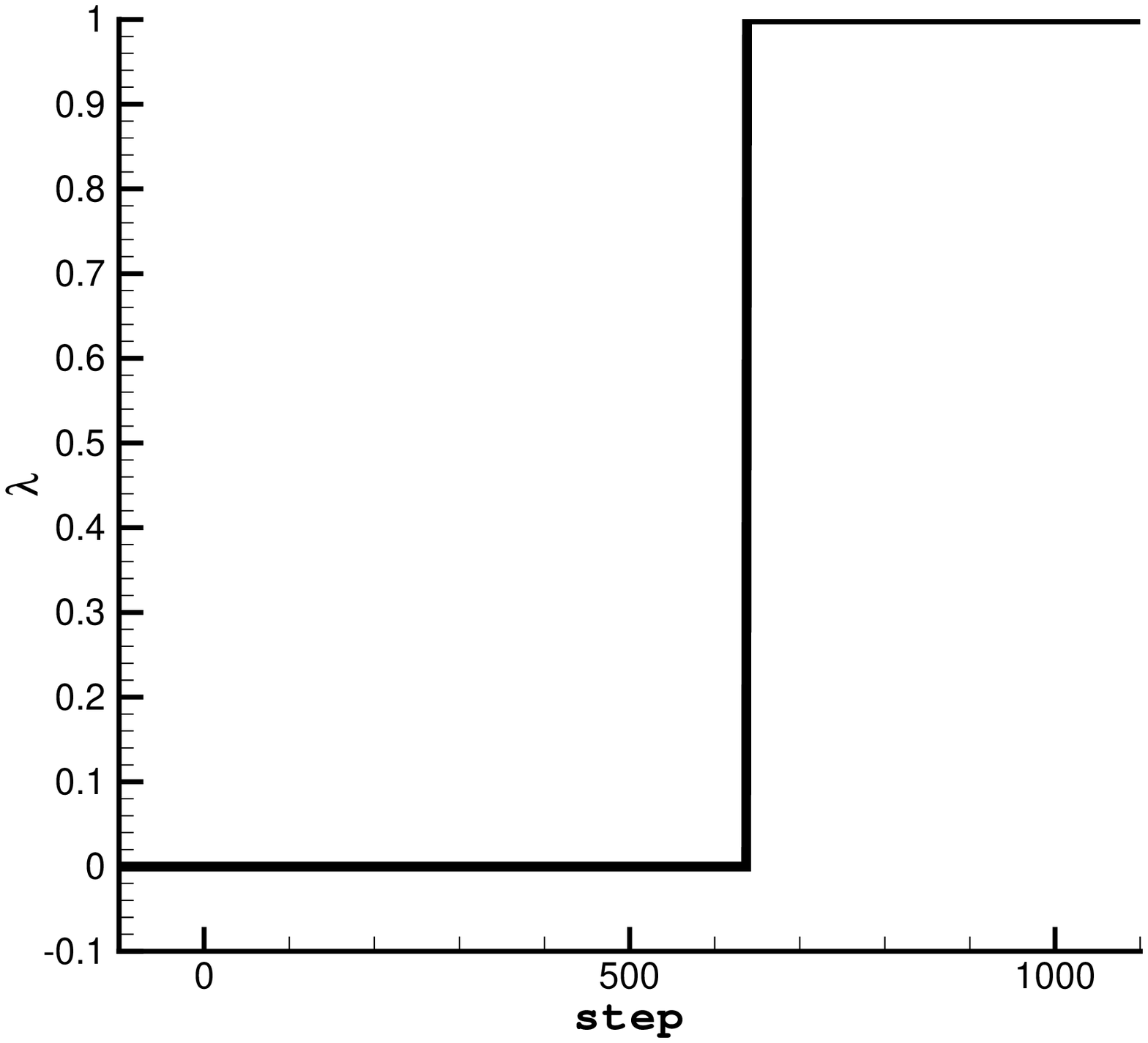}}
    \subfigure[]{\includegraphics[width=2.5in]{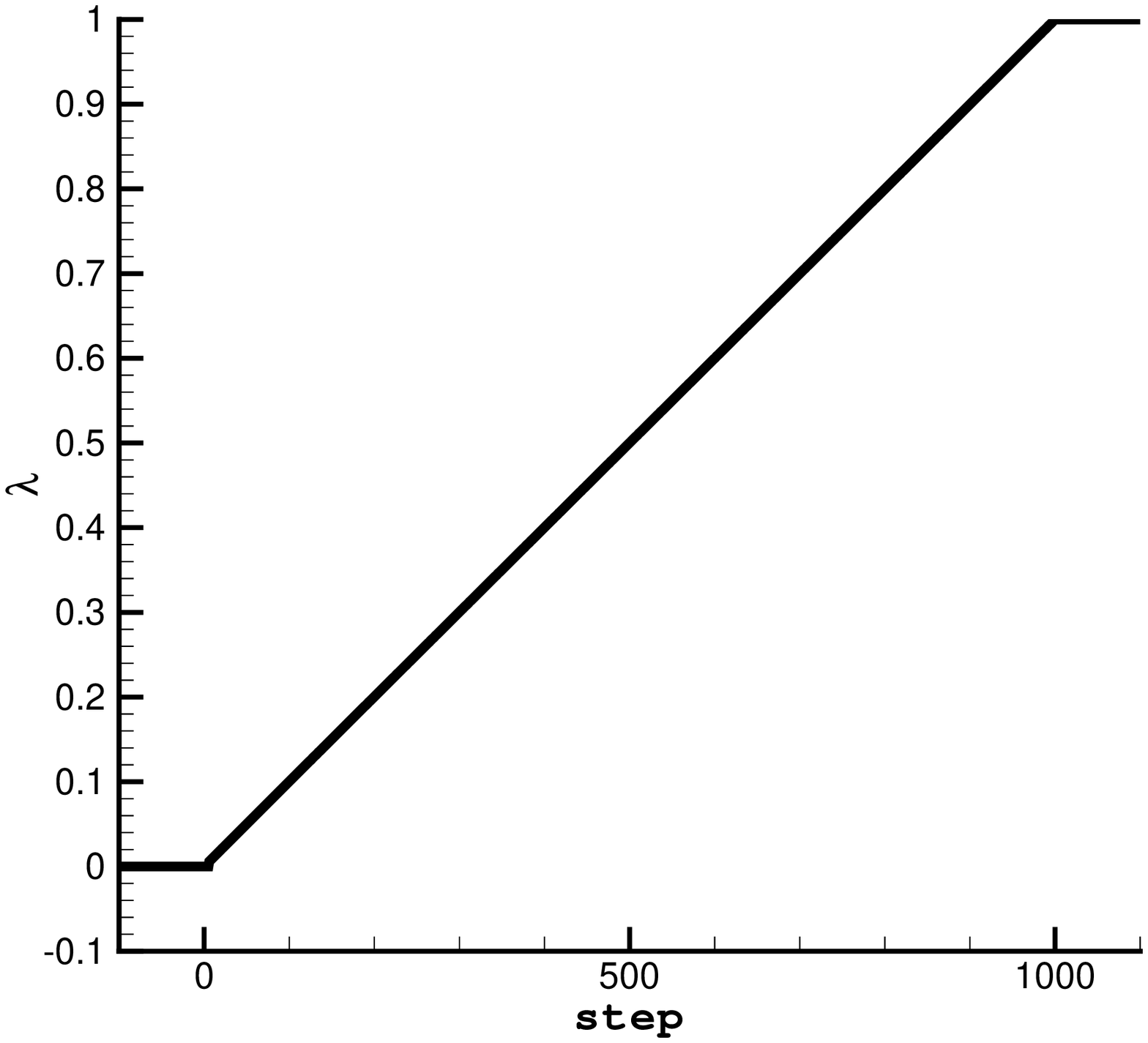}}
    \caption[]{Martensite-to-martensite transition example.
    Comparison of unrelaxed (left) and relaxed (right) energies,
    stresses, and volume fractions.} \label{fig:simplecase}
    \end{center}
\end{figure}

\begin{figure}[htbp]
    \begin{center}
    \subfigure[]{\includegraphics[width=3.15in]{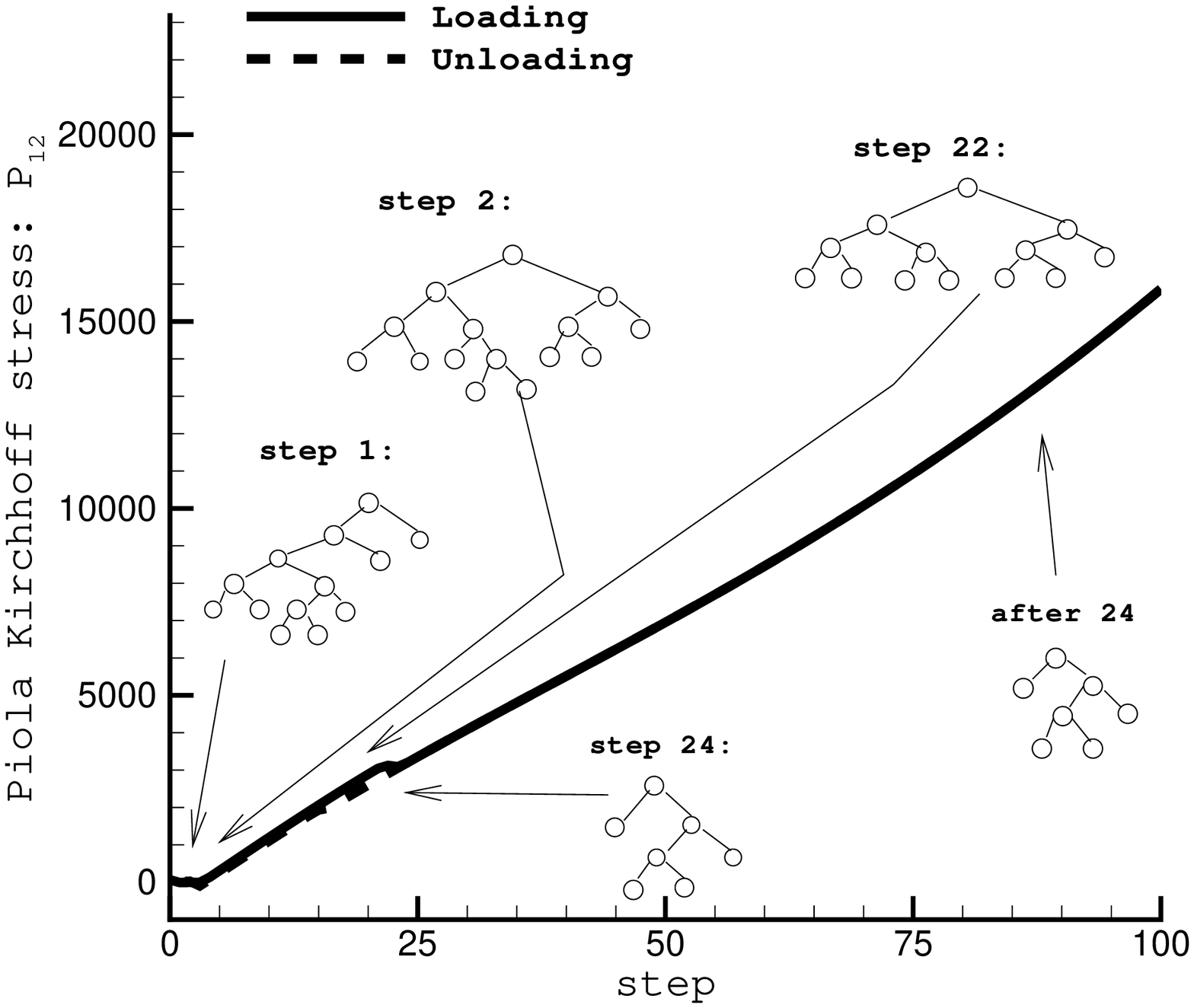}}
    \subfigure[]{\includegraphics[width=3.15in]{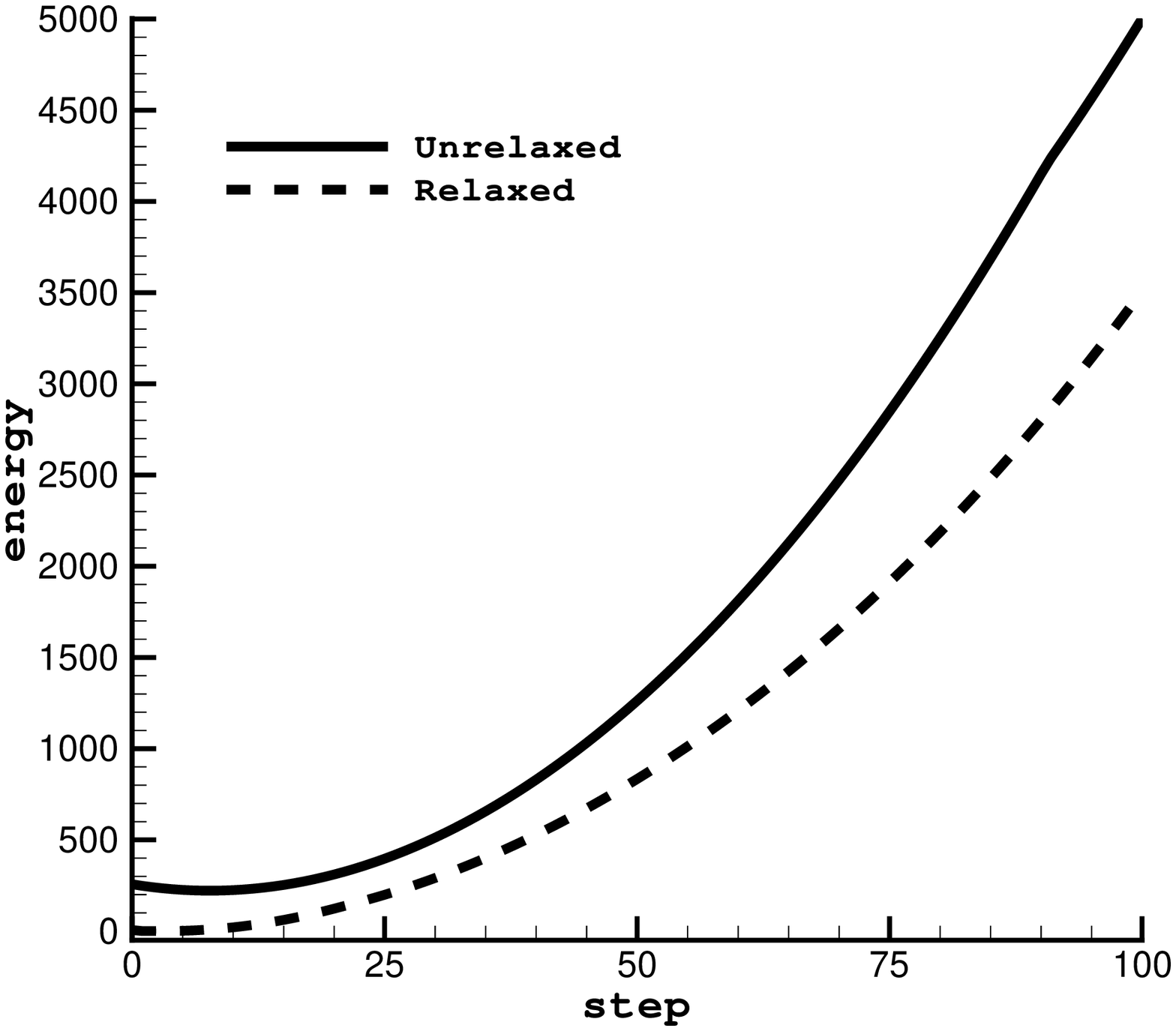}}
    \caption[]{Simple shear example. a) $P_{12}$ component of the
    first Piola-Kirchhoff stress tensor as a function of applied shear
    strain. The graph of the microstructures predicted by the relaxation
    algorithm are also shown inlaid. b) Unrelaxed and relaxed
    energies during loading.}\label{fig:stress-energ-gen}
  \end{center}
\end{figure}

\subsection{Martensite-martensite transition}

Our first test case concerns the macroscopic deformation process
\begin{equation}
\bF(t) = (1-t) \bU_1 + t \bU_2, \quad t \in [0,\,1]
\end{equation}
which takes the material from one variant of martensite to
another. Fig.~\ref{fig:simplecase} shows the evolution of the
energy, the component $P_{13}$ of the first Piola-Kirchhoff
stress, and the volume fraction $\lambda$ of $\bU_2$,
respectively, for the unrelaxed and relaxed cases. In this
calculations, the branching constructions employ the constrained
geometry as initial guess, as discussed in the foregoing.

The unrelaxed response shown in Fig.~\ref{fig:simplecase} exhibits
an abrupt transition from the initial to the final variant, as no
mixed states are allowed to develop during the deformation
process. By contrast, the relaxation algorithm results in the
development of a rank-one laminate immediately following the
onset of deformation. It should be carefully noted that the
material is allowed to develop laminates of arbitrary rank, and
that the persistency of a rank-one laminate is due to the fact
that both leaves are stable against branching. The computed volume
fraction $\lambda$ increases linearly from $0$ to $1$. As a
result of this evolving microstructure, the energy of the
material is fully relaxed, and the material remains unstressed.
In this example, the unloading response exactly traces in reverse
the loading response, and hence no hysteresis is recorded.

\subsection{Simple shear}

Our second test case concerns macroscopic simple shear on the
plane $(010)$ and in the direction $(100)$. The material is taken
to be initially undeformed, and the shear deformation is initially
increased from zero up to a maximum value and then decreased to
back zero. The calculations are carried out excluding the
austenitic well from the definition (\ref{eq:Multiwell}) of the
energy and considering the six martensitic wells only.

Fig.~\ref{fig:stress-energ-gen}b shows a comparison of the
computed unrelaxed and relaxed energies. By the exclusion of the
austenitic well the material is forced to develop a rank-five
laminate in its initial undeformed configuration. The graph of
this laminate, and of all laminates which subsequently arise
during deformation, is shown inlaid in
Fig.~\ref{fig:stress-energ-gen}a. As is evident from
Fig.~\ref{fig:stress-energ-gen}b, the relaxation algorithm
ostensibly succeeds at fully relaxing the energy of the material.
With increasing deformation, the computed microstructure
undergoes transitions to rank-four and three laminates. A first
rank-three laminate of order thirteen is first predicted which
subsequently simplifies to a rank-three laminate of order seven.
As a result of this microstructural evolution, the relaxed energy
remains well below the unrelaxed energy through the deformation,
Fig.~\ref{fig:stress-energ-gen}b. Upon unloading, the order-seven
rank-three laminate is maintained down to zero deformation,
suggesting that the initial microstructures computed during
loading are metastable. Indeed, the unloading stress-strain curve
lies below the loading one, resulting in a certain amount of
hysteresis, which suggests that the order-seven rank-three
laminate is indeed more efficient that the precursor
microstructures.

\section{Nonlocal extension}

The simple branching criterion (\ref{BranchingDrivingForce}),
which accounts for the energies of the variants only, neglects
any energy barriers as may opposed the transformation and may
lead to runaway refinement of the microstructure. In order to
limit branching and, additionally, to estimate the size of the
microstructure, we follow Ball and James \cite{ball.james2} and
take into consideration two additional sources of energy: the
energy of the twin boundaries, and the mismatch energy contained
within the boundary layers separating pairs of leaves. For
instance, Boullay {\it et al.} \cite{BoullaySchryversKohn2001},
and James {\it et al.} \cite{JamesKohnShield1995} have
investigated in detail the structure of branched needle
microstructures that develop within the misfit boundary layers,
e.~g., at the edge of a martensite laminate. Such level of detail
is well beyond the scope of this work. Our aim here is to derive
a rough estimate of the misfit energy amenable to a
straightforward calculation.

\begin{figure}[htbp]
\begin{center}
    \psfrag{F-}{$\bF^-$}
    \psfrag{F+}{$\bF^+$}
    \psfrag{l+}{$l^+$}
    \psfrag{l-}{$l^-$}
    \psfrag{lc}{$l_c$}
    \psfrag{N}{$\bN$}
    \psfrag{X3}{$X_3$}
    \psfrag{X1}{$X_1$}
    \psfrag{SamplingPoints}{Sampling Points}
    \psfrag{BoundaryLayer}{Boundary Layer}
    \includegraphics[width=3.5in, angle=0] {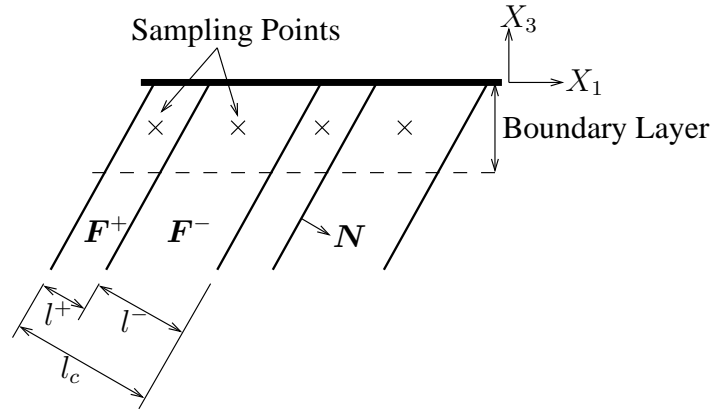}
    \caption{Schematic of interpolation boundary layer and scheme
    used to estimate misfit energy.} \label{fig:BL}
\end{center}
\end{figure}

One such simple estimate may be derived as follows. Begin by
enforcing `rigid-device' boundary conditions
\begin{equation}
\by^{\rm BL} = \by_0 + \bar{\bF}(\bx - \bx_0), \qquad x_3 = 0
\end{equation}
where $\bar{\bF}$ is the average deformation in the laminate,
$\bx_0$ is a material point within the reference configuration of
the laminate, and $\by_0 = \by(\bx_0)$ is its position on the
deformed configuration. This insulates the laminate from the
details of the adjacent deformation field in the regions $x_3 >
0$ and $x_3 < -l$. A simple interpolating deformation mapping is
then:
\begin{equation}
\by^{\rm BL}(\bx) = \by_0 - (\by(\bx) - \by_0) \frac{x_3}{\Delta}
+ \bar{\bF} (\bx - \bx_0) \left( 1 + \frac{x_3}{\Delta} \right),
\qquad - \Delta \leq x_3 \leq 0
\end{equation}
where $\by(\bx)$ is the deformation mapping of the laminate. The
boundary layer at $x_3 = -l$ can be given an identical treatment.
The corresponding deformation mapping is
\begin{equation}\label{FBL}
\bF^{\rm BL}(\bx) = - \bF(\bx) \frac{x_3}{\Delta} + \bar{\bF}
\left( 1 + \frac{x_3}{\Delta} \right) + \frac{1}{\Delta} [
\bar{\bF} (\bx - \bx_0) - (\by(\bx) - \by_0) ] \otimes \be_3
\end{equation}
We proceed to estimate the elastic energy of the region defined by
the intersection of each variant with the boundary layer by a
simple one-point quadrature rule. Let $\bx^\pm$ be a pair of
consecutive sampling points, Fig.~\ref{fig:BL}, chosen such that
$x^\pm_3 = - \Delta/2$, and select $\bx_0 = \bx^-$ for
simplicity. Then, it follows immediately from (\ref{FBL}) that
\begin{equation}
\bF^{{\rm BL} - } \equiv \bF^{\rm BL}(\bx^-) = \frac{1}{2} ( \bF^-
+ \bar{\bF})
\end{equation}
Likewise,
\begin{equation}
\bF^{{\rm BL} + } \equiv \bF^{\rm BL}(\bx^+) = \frac{1}{2} ( \bF^+
+ \bar{\bF}) + [ \bar{\bF} (\bx^+ - \bx^-) - (\by^+ - \by^-) ]
\otimes \be_3
\end{equation}
But, since $\by(\bx)$ is piecewise linear, it follows that
\begin{equation}
\by^+ - \by^- = \bF^- (\lambda^- (\bx^+ - \bx^-)) + \bF^+
(\lambda^+ (\bx^+ - \bx^-)) = \bar{\bF} (\bx^+ - \bx^-)
\end{equation}
and
\begin{equation}
\bF^{{\rm BL} + } = \frac{1}{2} ( \bF^+ + \bar{\bF})
\end{equation}
Collecting the above results we finally have
\begin{equation}\label{FBLSamplingPoints}
\bF^{{\rm BL} \pm} = \frac{1}{2} ( \bF^\pm + \bar{\bF})
\end{equation}
Within this approximation, the misift boundary-layer energy
density finally evaluates to
\begin{equation}\label{WBL}
W^{\rm BL} = \lambda^- [ W( \bF^{{\rm BL}-} ) - W( \bF^- )] +
\lambda^+ [ W( \bF^{{\rm BL}+} ) - W( \bF^+ )]
\end{equation}
which furnishes a remarkably simple (though rough) estimate. We
note that
\begin{equation}
\bF^{{\rm BL} + } - \bF^{{\rm BL} - } = \frac{1}{2} (\bF^+ -
\bF^-) = \frac{1}{2} \ba \otimes \bN
\end{equation}
and
\begin{equation}
\lambda^- \bF^{{\rm BL} - } + \lambda^+ \bF^{{\rm BL} + } =
\bar{\bF}
\end{equation}
Thus, the deformations $\bF^{{\rm BL} \pm}$ are rank-one
compatible and match the average deformation of the laminate.
Since the deformations $\bF^\pm$ minimize the energy of the
laminate among all rank-one laminates with average deformation
$\bar{\bF}$, it follows that $W^{\rm BL} \geq 0$, i.~e., $W^{\rm
BL}$ does indeed represents an \emph{excess energy}.

For simplicity, we assume that the twin-boundary energy $\Gamma$
per unit area is a constant independent of the deformation of the
variants. Combining the preceding estimates, it follows that
total excess or \emph{nonlocal} energy due to the twin boundaries
and the misfit boundary layers contained within a region of the
laminate of dimensions $L \times L \times l$ is:
\begin{equation}
E^{\rm NL} = L^2 l \left\{ \frac{\Gamma}{l^c} + \frac{ 2 \Delta
}{l} W^{\rm BL} \right\} \label{fnNonLocal}
\end{equation}
Taking $\Delta = l^c/2$, for definiteness, this expression reduces
to
\begin{equation}
E^{\rm NL} = L^2 l \left\{ \frac{\Gamma}{l^c} + \frac{ l^c }{l}
W^{\rm BL} \right\} \label{fnNonLocal2}
\end{equation}
This excess energy may now be minimized with respect to $l^c$,
with the result:
\begin{equation}\label{l}
l^c = \sqrt{\frac{\Gamma l}{W^{\rm BL}}}
\end{equation}
which affords an estimate of $l^c$. The corresponding minimum
excess energy per unit volume is
\begin{equation}
W^{\rm NL} \equiv \frac{E^{\rm NL}}{L^2 l} = 2 \sqrt{\frac{\Gamma
W^{\rm BL} }{l}}
\end{equation}
We note that this excess energy grows as $l^{-1/2}$, which tends
to suppress microstructural refinement. In calculations, we
interpret the excess energy density $W^{\rm NL}$ as an energy
barrier for branching. Consideration of this energy barrier has
the effect of reducing the local branching driving force
(\ref{BranchingDrivingForce}) to
\begin{equation}\label{BranchingDrivingForce2}
f_l^n = W(\bF_l^n) - R_1W(\bF_l^n) - W^{\rm NL}(\bF_l^n)
\end{equation}
which effectively introduces a lower cut-off for the size of the
microstructure and eliminates the possibility of runaway
microstructural refinement.

\section{Application to the finite-element simulation of
indentation in \CAN}

The sequential lamination algorithm developed in the foregoing
may conveniently be taken as a basis for multiscale simulation in
situations in which there is a strict separation of scales: a
macroscopic scale characterized by slowly-varying smooth fields;
and a much smaller scale commensurate with the size of the
evolving microstructure. As remarked by several authors
\cite{chipot.kinderlehrer, luskin, carstensen.plechac, desimone},
these problems may be solved effectively by pushing the
microstructure to the \emph{sub-grid} scale, while solving the
well-posed relaxed problem on the computational grid.

\begin{figure}[htbp]
  \begin{center}
    \includegraphics[width=8cm]{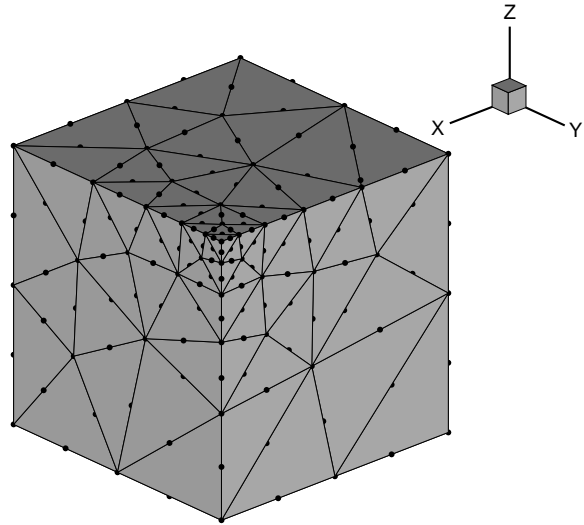}
    \caption{\em Computational domain and finite element mesh.}
    \label{fig:init-mesh}
  \end{center}
\end{figure}

In this section we present selected examples of application of
this multiscale approach in which the macroscopic problem is
solved by the finite-element method, while the effective behavior
is computed, simultaneously with the macroscopic solution, at the
Gauss-point level using the sequential lamination algorithm
developed in the foregoing. The particular problem considered
concerns the quasistatic normal indentation of a \CAN shape-memory
alloy by a spherical indenter. The domain of analysis and the
computational mesh are shown in Fig.~\ref{fig:init-mesh}. The
analysis is reduced to one quarter of the entire domain for
simplicity. In particular, solutions exhibiting broken symmetry
are ruled out by the analysis. The size of the computational
domain is 20mm $\times$ 20mm $\times$ 20mm. The radius of the
indenter is 15mm. The specimen is fully supported over its entire
base, and the remainder of its boundary is free of tractions. The
computational mesh contains $254$ nodes and $105$ ten-node
quadratic tetrahedral elements. Contact between the indenter and
the specimen is assumed to be frictionless and is enforced by a
penalty method \cite{pandolfi}. To ensure that the jacobian $J$ of
the deformation remains positive in all variants at all times, the
energy of each well is augmented by a term of the form \cite{neff}
\begin{equation}
W^{\rm vol}(J) = \left\{
\begin{matrix}
C (J^2 + J^{-2} - 2)^2, & J < 1 \\
0, & \text{otherwise}
\end{matrix}
\right.
\end{equation}
where $C$ is a constant chosen sufficiently small to minimize the
effect on the total energy. By design, $W^{\rm vol}(J)$ and its
first and second derivatives vanish at $J=1$. In addition, the
twin-boundary and misfit energies are accounted for as part of the
branching criterion as a means of introducing a lower cutoff for
the laminate size and preventing runaway microstructural
refinement. In all calculations, the twin-boundary energy per unit
area $\Gamma$ is set to 1 J/m${}^2$. The maximum size of the
laminate at a particular Gauss point is set to the element size,
in keeping with the assumption that the laminate accounts for
sub-grid structure in the solution only, and that coarser
structures are adequately resolved by the mesh.

The finite-element solution is obtained by dynamic relaxation
followed by a preconditioned conjugate-gradient iteration
\cite{shewchuk}. The high level of concurrency in the
constitutive calculations was exploited via an MPI-based parallel
implementation \cite{pacheco} on the ASCI Blue multiprocessing
computer. Performance studies showed excellent load balancing and
scalability.

\begin{center}
\begin{figure}
    \subfigure[]{\includegraphics[width=8.50cm]{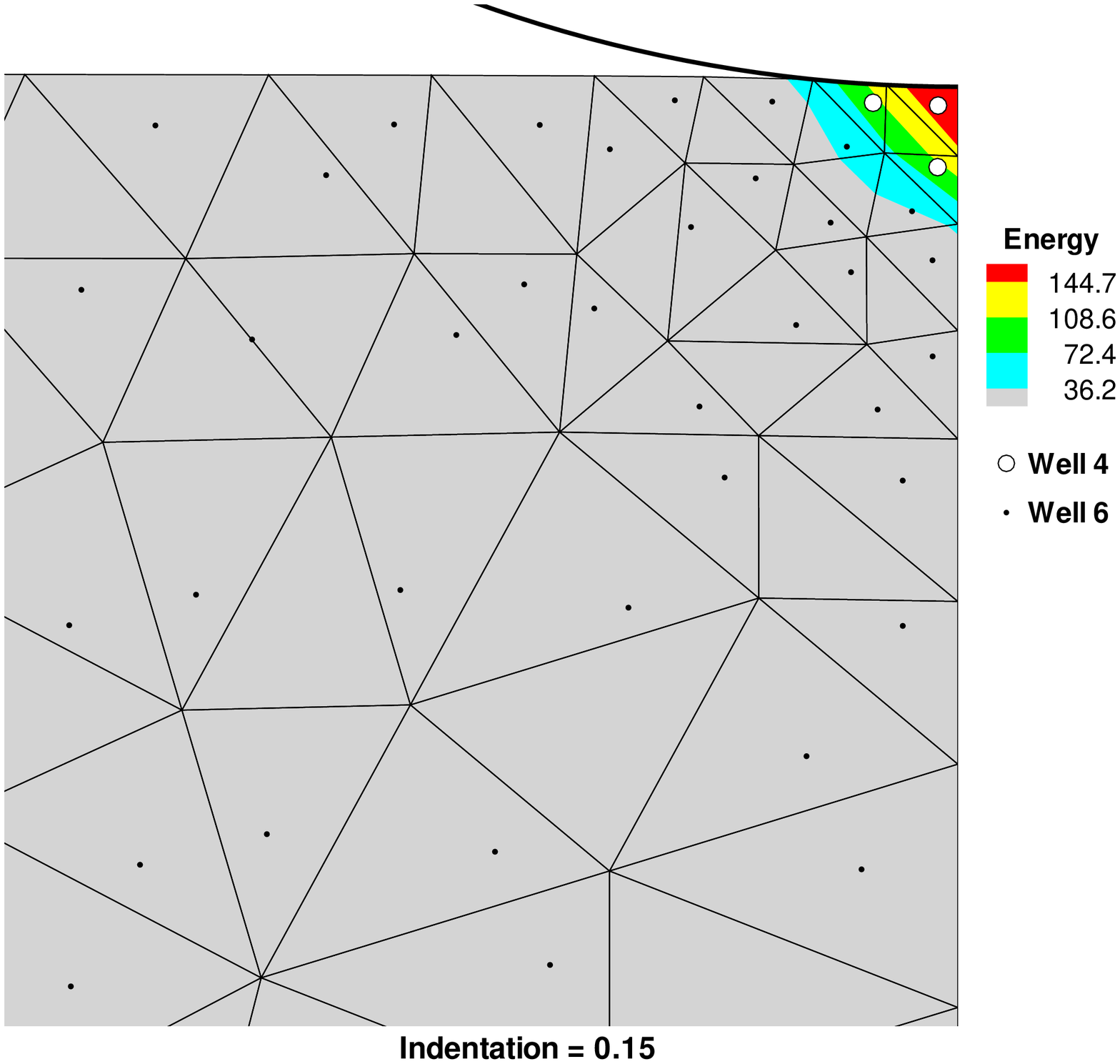}}
    \subfigure[]{\includegraphics[width=8.50cm]{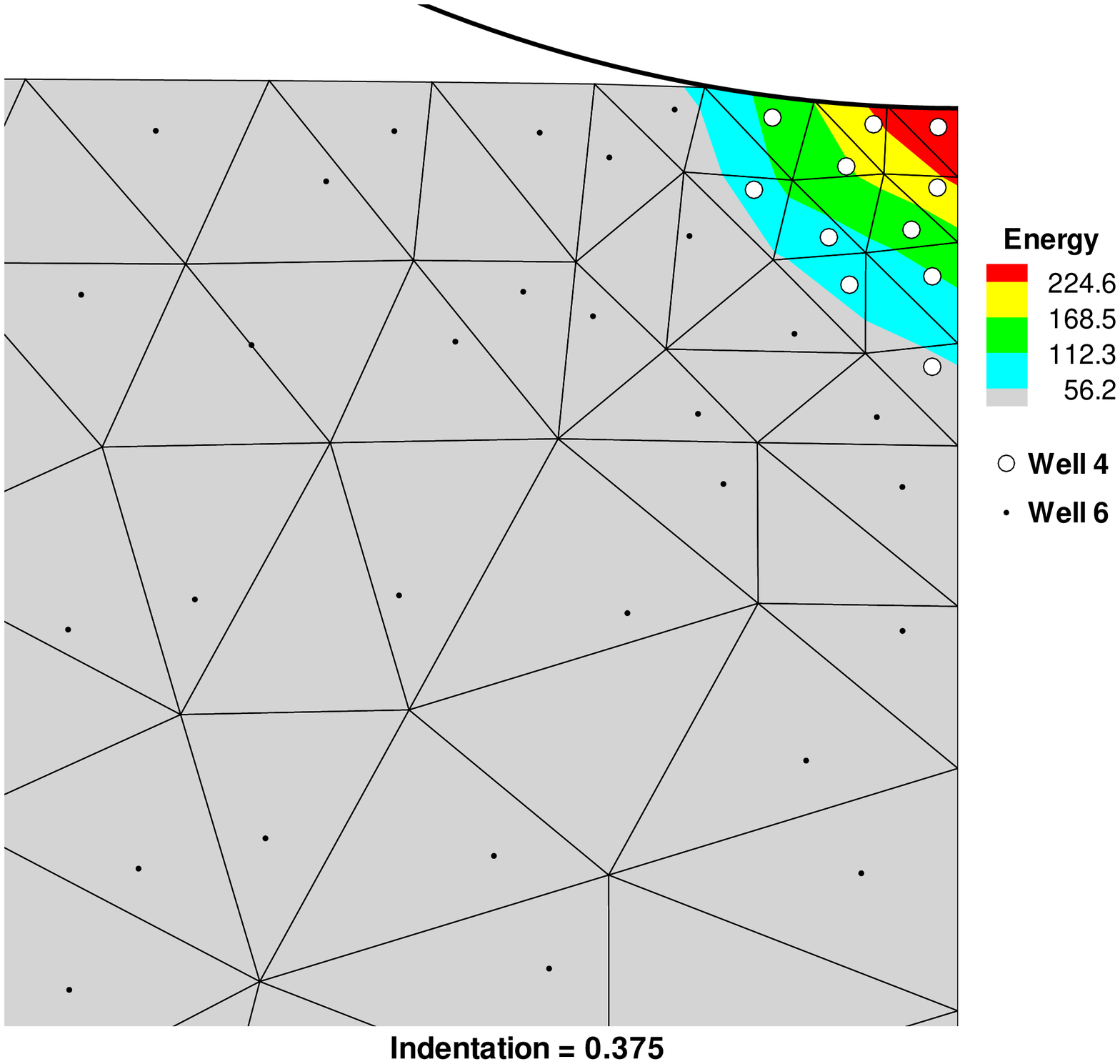}}
    \caption[]{Cross-sections and energy-density contours for
    two unrelaxed solutions at depths of indentation: a) 0.150 mm,
    and b) 0.375 mm. The symbols designate the energy well which is
    activated at each Gauss point of the mesh.}
\label{fig:mesh-unrelaxed}
\end{figure}
\end{center}

\begin{center}
\begin{figure}
    \subfigure[]{\includegraphics[width=8.cm]{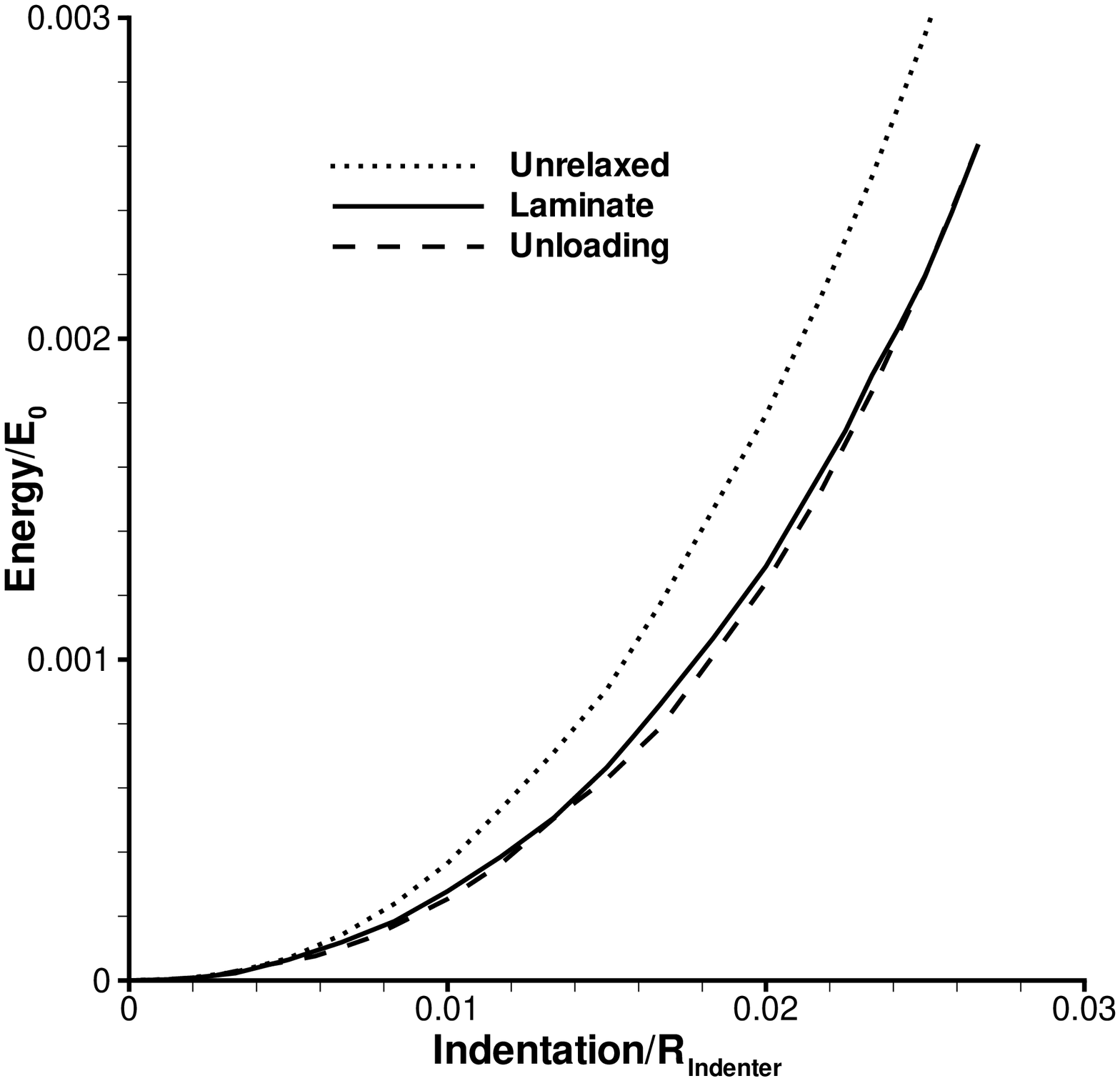}}
    \subfigure[]{\includegraphics[width=8.cm]{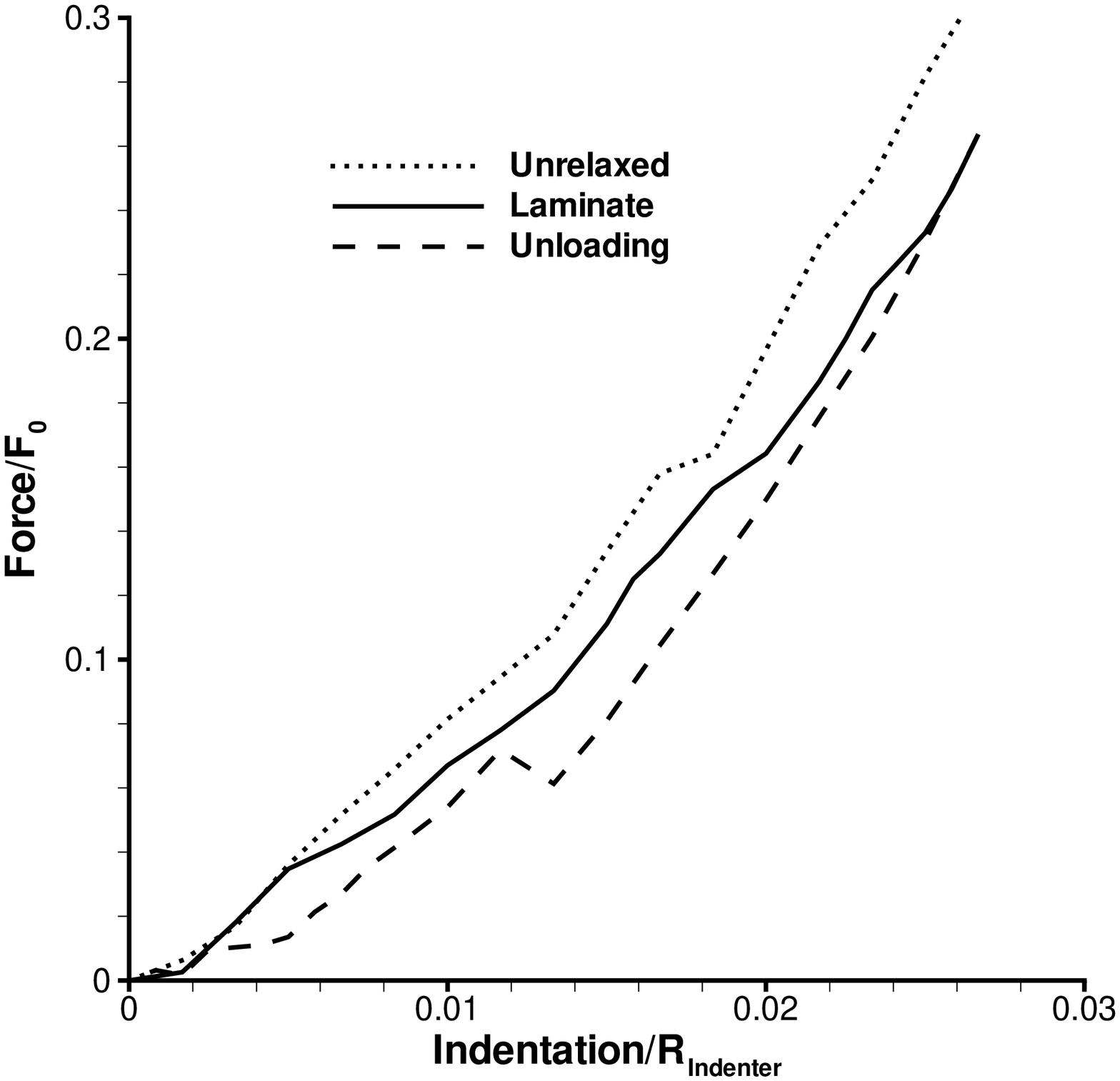}}
    \caption[]{Normalized total energy {\it vs.} normalized depth
    of indentation; b) Normalized indentation force {\it vs.}
    normalized depth of indentation.}
\label{fig:force-energy}
\end{figure}
\end{center}

Fig.~\ref{fig:mesh-unrelaxed} shows the \emph{unrelaxed} deformed
configurations, and the corresponding distribution of active energy
wells at the Gauss points of the mesh, at depths of indentation of
0.150 and 0.375 mm. As is evident from this figure, two energy wells
become active during indentation. The transformed zone under the
indenter grows with depth of indentation, but the fineness of the
variant arrangement is severely limited by the mesh
size. Correspondingly, the total energies and indentation forces
recorded during indentation are comparatively high,
Fig.~\ref{fig:force-energy}. In this figure the energy has been
normalized by $E_0=V_0C_{11}^{Austenite}$, where $V_0$ is the volume
of the undeformed specimen, while the force has been normalized by
$F_0=E_0/R_{Indenter}$.

\begin{center}
\begin{figure}
    \subfigure[]{\includegraphics[width=8.cm]
        {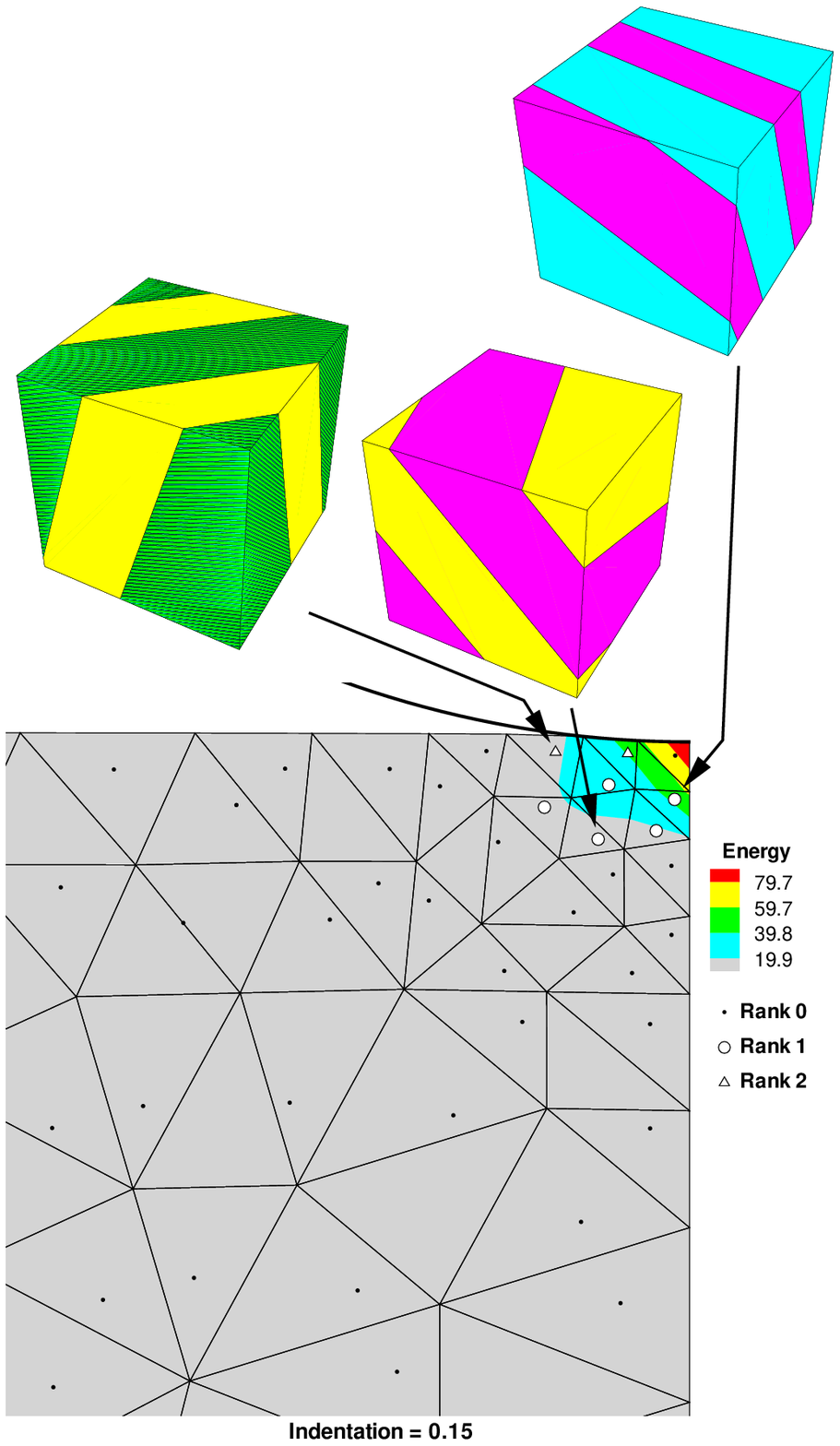}}
    \subfigure[]{\includegraphics[width=8.cm]
        {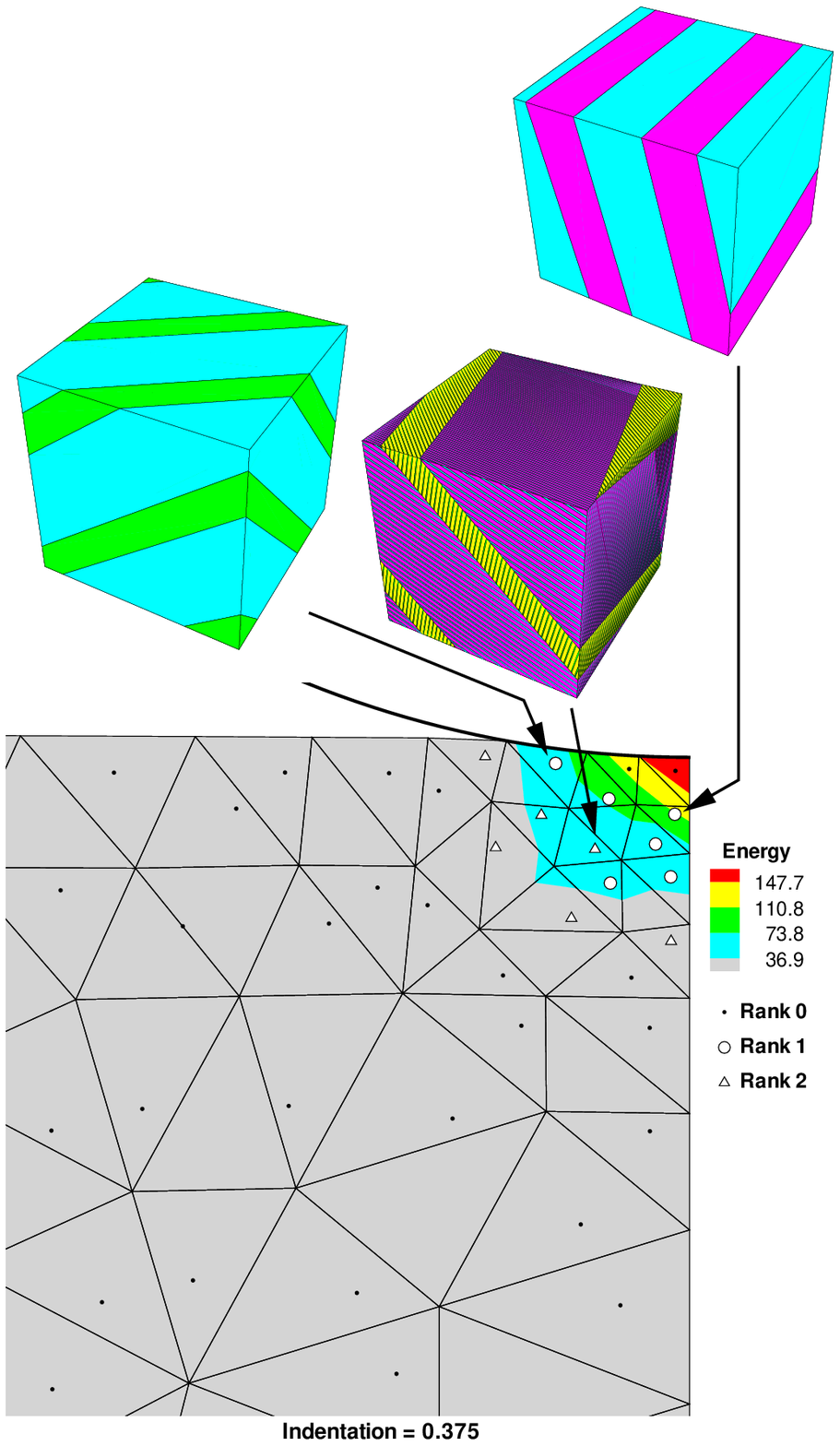}}
    \caption[]{Cross-section and energy-density contours for
    relaxed solution at an indentation depth of: a) 0.150 mm, and b)
    0.375 mm. The symbols indicate the rank of the microstructure at
    the Gauss  points. The insets depict the geometry of the
    microstructure at the indicated sampling points, with each
    color representing an individual well, and are of identical
    size oriented such that the left face corresponds to the
    cross-section plane.} \label{fig:lamplots}
\end{figure}
\end{center}

The \emph{relaxed} solution obtained using the sequential
lamination algorithm differs markedly from the unrelaxed solution
just described, Fig.~\ref{fig:lamplots}. Thus, the relaxed
deformation field is accompanied by the development of
well-defined microstructures at the subgrid level. Some of the
laminates generated by the sequential lamination algorithm are
quite complex, reaching rank two. Of note is the appearance of a
de-twinned zone directly under the indenter. The effect of
relaxation on the total energy and indentation force is quite
marked, Fig.~\ref{fig:force-energy}, with the relaxed values lying
well below the unrelaxed ones. Unloading exhibits the
path-dependent nature of the algorithm, with the microstructure
established at maximum load remaining in place during much of the
unloading process, which in turn results in a soft response. The
fineness of the microstructure is somewhat overpredicted by the
calculations, with some of the variants attaining sub-micron
thicknesses. In view of (\ref{l}), this excessive refinement may
owe to a low value of the twin-boundary energy $\Gamma$, or to an
overestimation of the misfit energy $W^{\rm BL}$, or both.

\section{Summary and Concluding Remarks}

We have presented a practical algorithm for partially relaxing
multiwell energy densities. The algorithm is based on sequential
lamination, but it is constrained in such a way that successive
microstructures occurring along a deformation path are close to
each other in a certain sense: the new microstructure should be
reachable from the preceding one by a combination of branching and
pruning operations. All microstructures generated by the algorithm
are in static and configurational equilibrium. In particular, we
optimize all the interface orientations and variant volume
fractions, with the result that all configurational forces and
torques are in equilibrium. We additionally allow the variants to
be arbitrarily stressed and enforce traction equilibrium across
all interfaces. Owing to the continuity constrained imposed upon
the microstructural evolution, the predicted material behavior may
be path-dependent and exhibit hysteresis.

In cases in which there is a strict separation of micro and
macrostructural lengthscales, the proposed relaxation algorithm
may effectively be integrated into macroscopic finite-element
calculations at the subgrid level. We have demonstrated this
aspect of the algorithm by means of a numerical example concerned
with the indentation of an \CAN shape memory alloy
\cite{chu.james} by a spherical indenter. The calculations
illustrate the ability of the algorithm to generate complex
microstructures, resulting in force-depth of indentation curves
considerably softer than otherwise obtained by direct energy
minimization.

Several improvements of the present approach immediately suggest
themselves. The relaxation algorithm, in its present form, does
not account for energy barriers for branching. Thus, a significant
improvement over this model would be to permit, with probability
less than one, transitions requiring an energy barrier to be
overcome, e.~g., in the spirit of transition state theory and
kinetic Monte Carlo methods. This extension would require a
careful and detailed identification of all the paths by which
branching may take place, and the attendant energy barriers.
Another significant improvement would be to relax the
configurational force equilibrium constraint and replace it by a
kinetic relation governing interfacial motion
\cite{abeyaratne.chu.james, AbeyaratneKnowles1997, bhattacharya}.
Kinetic relations of this form can effectively be integrated into
the variational principle with the aid of time discretization
\cite{ortiz.repetto, ortiz.stainier}.

\bigskip
\section*{Acknowledgments}

We are grateful for support provided by the US Department of
Energy through Caltech's ASCI/ASAP Center for the Simulation of
the Dynamic Behavior of Solids; and for support provided by the
AFOSR through Brown's MURI for the Design of Materials by
Computation. MF is also grateful for support provided by the DOE
and Krell Institute through a Computational Science Graduate
Fellowship.

\nocite{*}
\bibliographystyle{plain}

\bibliography{biblio}

\end{document}